\documentclass[paper]{JHEP3} 

\JHEPspecialurl{http://jhep.sissa.it/JOURNAL/JHEP3.tar.gz}

\usepackage{epsfig,multicol}

\newcommand\fverb{\setbox\pippobox=\hbox\bgroup\verb}
\newcommand\fverbdo{\egroup\medskip\noindent%
			\fbox{\unhbox\pippobox}\ }
\newcommand\fverbit{\egroup\item[\fbox{\unhbox\pippobox}]}
\newcommand {\beq}{\begin{equation}}
\newcommand {\eeq}{\end{equation}}
\newcommand {\bea}{\begin{eqnarray}}
\newcommand {\eea}{\end{eqnarray}}
\newcommand {\nn}{\nonumber}
\newcommand {\tr}{{\rm tr\,}}

\newcommand {\dd}{\mbox{d}}

\newcommand {\del}{\partial}

\newcommand {\bA}{{\tt A}}
\newcommand {\bB}{{\tt B}}
\newcommand {\bC}{{\tt C}}
\newcommand {\limit}{\rightarrow}

\newbox\pippobox

\title{Various Super Yang-Mills Theories 
with Exact Supersymmetry on the Lattice}
\author{Fumihiko Sugino\\
Okayama Institute for Quantum Physics\\
Kyoyama 1-9-1, 
Okayama 700-0015, Japan\\
E-mail: \email{fumihiko\_sugino@pref.okayama.jp}}

\preprint{OIQP-04-2 \\ \heplat{0410035}}	

\abstract{We continue to construct lattice super Yang-Mills theories 
along the line discussed in the previous papers \cite{sugino, sugino2}. 
In our construction of ${\cal N}=2, \, 4$ theories in four dimensions, 
the problem of degenerate vacua seen in \cite{sugino} 
is resolved 
by extending some fields and soaking up would-be zero-modes in the continuum limit, 
while in the weak coupling expansion 
some surplus modes appear both in bosonic and fermionic sectors reflecting the exact supersymmetry. 
A slight modification to the models is made such that all the surplus modes are 
eliminated in two- and three-dimensional models obtained by dimensional reduction thereof.     
${\cal N}=4, \, 8$ models in three dimensions need fine-tuning of three and one 
parameters respectively to obtain the desired continuum theories, while 
two-dimensional models with ${\cal N}=4, \, 8$ do not require any fine-tuning. }


\begin{document} 


\section{Introduction}
In this paper, we continue to discuss on a lattice formulation of supersymmetric Yang-Mills (SYM) 
theories with extended supersymmetries along the line of the papers \cite{sugino,sugino2}. 
The formulation is based on `topological field theory (TFT) formulation' \cite{tft} and 
`balanced  topological field theory (BTFT) formulation' \cite{btft} of ${\cal N}=2, \, 4$ SYM theories, 
respectively\footnote{See refs. \cite{catterall} 
for related constructions of lattice models for theories without gauge symmetry.}. 
It exactly realizes one or two supercharges on the lattice which 
can be regarded as a fermionic internal symmetry. 
See refs. \cite{kikukawa-nakayama}-\cite{kaplan_related} 
for some recent works on other approaches to lattice 
supersymmetry. (Also, for recent reviews, 
see \cite{review}.)

In ref. \cite{sugino}, we discussed on a general aspect of the formulation, and encountered a problem of 
degenerate vacua of the same kind as in \cite{elitzur}. 
There we resolved it introducing supersymmetry breaking 
terms which become irrelevant in the continuum limit. 
In \cite{sugino2}, focusing on two-dimensional lattice 
SYM theories with ${\cal N}=2, \, 4$, we have presented modifications 
to resolve the degenerated vacua {\it without affecting the exact supersymmetry}. 
It has been shown that in the continuum limit 
the two-dimensional models flow to the desired continuum theories possessing 
full supersymmetry and rotational invariance without fine-tuning. 
 
Here, we first consider lattice models for four-dimensional SYM theories with ${\cal N}=2, \, 4$ 
and their dimensionally reduced versions. 
The problem of the vacuum degeneracy is resolved in four dimensions 
by extending some fields and soaking up would-be zero-modes in the continuum limit, 
while surplus zero-modes of the kinetic terms 
with nonzero momenta appear after the weak field expansion of the gauge link variables. 
As a manifestation of the exact supersymmetry, such modes exist 
in both of bosonic and fermionic sectors with the equal multiplicity. 
Because each of the surplus modes has smaller degrees of freedom than the ordinary 
zero-mode with zero-momentum, strictly speaking 
they are not the so-called doublers which have the degrees of freedom equivalent 
to those of the zero-mode with zero-momentum. 
Their appearence is considered to reflect the fact that our models do not exactly realize a lattice 
counter part of the topological term $\int \tr F \wedge F$. 
It is not a topological quantity on the lattice in the sense that an arbitrary small variation of 
the quantity does not always vanish, and it has some surplus dynamical modes, which we are observing as 
the surplus modes. 
In this paper, we will not pursue more on the four-dimensional models. 
Instead, we make use of them 
as an intermediate step to construct two- and three-dimensional models 
{\it without surplus modes}. 
This is a main issue of this paper. 
The lower-dimensional models have ${\cal N}=4, \, 8$ supersymmetries 
in two and three dimensions, 
and they are obtained by dimensional reduction from the four-dimensional models 
after a slight modification done. 
The renormalization arguments 
based on the dimensional analysis and symmetries of the lattice action show that three-dimensional 
${\cal N}=4, \, 8$ cases lead the target continuum theories with tuning three and one parameters 
respectively, while the two-dimensional cases need no fine-tuning.   
  
This paper is organized as follows. 
In section 2, we discuss a naive model for 
four-dimensional ${\cal N}=2$ SYM theory as well as its slightly modified version, 
and observe surplus zero-modes appearing 
in the kinetic terms. 
In section 3, by considering dimensional reduction of the slightly modified four-dimensional model, 
we construct lattice models free from the surplus modes defined in two and three dimensions. 
Also, the renormalization arguments are presented near the continuum limit. 
Section 4 is devoted to both of a naive lattice model and its slightly modified one 
for ${\cal N}=4$ SYM theory in 
four dimensions. In section 5, ${\cal N}=8$ SYM models on the two and three dimensional lattices 
are presented 
and their renormalization properties are examined. 
We summarize the result obtained so far 
and discuss some future directions in section 6. Discussions on the resolution of the vacuum degeneracy 
are presented in appendices A and B. Appendix C gives 
the explicit form of the matrices 
$\gamma_{\mu}$, $P_{\mu}$, $P_{\mu}'$ 
appearing in the fermion kinetic terms of various models. 

Throughout this paper, we consider the gauge group $G= \mbox{SU}(N)$ and 
the $d$-dimensional hypercubic lattice ${\bf Z}^d$ $(d=2, \, 3, \, 4)$.  
Gauge fields are expressed as compact unitary variables 
\beq
U_{\mu}(x)= e^{iaA_{\mu}(x)}
\label{unitary}
\eeq
on the link $(x, x+\hat{\mu})$. 
`$a$' stands for the lattice spacing, 
and $x\in {\bf Z}^d$ the lattice site. 
Other fields, put on the sites, are traceless hermitian matrices expanded by the SU($N$) generators
(the complex scalars $\phi$, $\bar{\phi}$ are complexified versions of them
). We extend some fields to hermitian matrices with nonvanishing trace parts, 
which are expressed by putting the hat ($\hat{~~}$).

\setcounter{equation}{0}
\section{Naive 4D ${\cal N}=2$ Lattice Model and Its Slight Modification}
\label{sec:4DN2}
We consider a naive lattice model for four-dimensional ${\cal N}=2$ SYM theory 
as well as its slightly modified version, which are  
almost same as the model presented in 
ref. \cite{sugino}. 
Notations are same as in \cite{sugino}.  

\subsection{Naive Lattice Model for 4D ${\cal N}=2$}
We start with the lattice action for four-dimensional ${\cal N}=2$:  
\bea
S^{{\rm LAT}}_{4D{\cal N}=2} & = & Q\frac{1}{2g_0^2}\sum_x \, \tr\left[ 
\frac14 \eta(x)\, [\phi(x), \,\bar{\phi}(x)] -i\vec{\chi}(x)\cdot(\vec{\Phi}(x) 
+ \Delta\vec{\Phi}(x))
+\vec{\chi}(x)\cdot\vec{H}(x)\right. \nn \\
 & & \hspace{2cm}\left. \frac{}{} 
+i\sum_{\mu=1}^4\psi_{\mu}(x)\left(\bar{\phi}(x) - 
U_{\mu}(x)\bar{\phi}(x+\hat{\mu})U_{\mu}(x)^{\dagger}\right)\right], 
\label{4DN2_Sn1}
\eea
where $\vec{H}$, $\vec{\chi}$, $\vec{\Phi}$, $\Delta\vec{\Phi}$ are three-component vectors.  
In terms of the plaquette variables 
\beq
U_{\mu\nu}(x) \equiv U_{\mu}(x) U_{\nu}(x+\hat{\mu}) 
U_{\mu}(x+\hat{\nu})^{\dagger} U_{\nu}(x)^{\dagger}, 
\eeq
$\vec{\Phi}(x)$ and $\Delta\vec{\Phi}(x)$ are expressed as 
\bea
\Phi_1(x) & = & -i\left[U_{1 4}(x) -U_{41}(x) 
+U_{23}(x) - U_{32}(x)\right], \nn \\
\Phi_2(x) & = & -i\left[U_{2 4}(x) -U_{42}(x) 
+U_{31}(x) - U_{13}(x)\right], \nn \\
\Phi_3(x) & = & -i\left[U_{34}(x) -U_{43}(x) 
+U_{12}(x) - U_{21}(x)\right], 
\nn \\
\Delta\Phi_1(x) & = & -r\left[W_{1 4}(x) 
+ W_{23}(x)\right], \nn \\
\Delta\Phi_2(x) & = & -r\left[W_{2 4}(x) 
+ W_{31}(x)\right], \nn \\
\Delta\Phi_3(x) & = & -r\left[W_{3 4}(x) 
+ W_{12}(x)\right], 
\label{DPhin1}
\eea
with 
\beq
W_{\mu\nu}(x) \equiv 2-U_{\mu\nu}(x) - U_{\nu\mu}(x).
\label{def_W}
\eeq
$\phi$, $\bar{\phi}$ represent complex scalar fields, and $\psi_{\mu}$, $\vec{\chi}$, $\eta$ fermions. 
Also, $\vec{H}$ are auxiliary fields. 
The nilpotent supersymmetry $Q$: 
\beq
Q^2 = (\mbox{infinitesimal gauge transformation with the parameter } \phi)
\label{Q_nilpotent}
\eeq
acts to the lattice fields as 
\bea
 & & QU_{\mu}(x) = i\psi_{\mu}(x) U_{\mu}(x), \nn \\
 & & Q\psi_{\mu}(x) = i\psi_{\mu}(x)\psi_{\mu}(x) 
    -i\left(\phi(x) - U_{\mu}(x)\phi(x+\hat{\mu})U_{\mu}(x)^{\dagger}\right),
  \nn \\
 & & Q\phi(x) = 0,     \nn \\
 & & Q\vec{\chi}(x) = \vec{H}(x), \quad 
           Q\vec{H}(x) = [\phi(x), \,\vec{\chi}(x)], \nn \\
 & & Q\bar{\phi}(x) = \eta(x), \quad  Q\eta(x) = [\phi(x), \,\bar{\phi}(x)]. 
\label{Q_lattice}
\eea
The action has the $Q$ exact form based on the `TFT formulation' \cite{tft} 
of four-dimensional ${\cal N}=2$ 
SYM discussed in ref. \cite{sugino}. 
This $Q$, preserved at the lattice level, is one of eight supercharges of four-dimensional 
${\cal N}=2$ supersymmetry. 
The action has the global U$(1)_R$ internal symmetry:  
\bea
 & & \psi_{\mu}(x)\limit e^{i\Lambda}\psi_{\mu}(x), \quad 
\vec{\chi}(x)\limit e^{-i\Lambda}\vec{\chi}(x), \quad \eta(x) \limit e^{-i\Lambda}\eta(x), \nn \\
 & & \phi(x) \limit e^{i2\Lambda}\phi(x), \quad \bar{\phi}(x) \limit e^{-i2\Lambda}\bar{\phi}(x)
\label{U1R}
\eea
with other fields intact. 

Note that we employ dimensionless variables on the lattice, and thus 
various quantities are of the following orders:  
\bea
& & \psi_{\mu}(x), \vec{\chi}(x), \eta(x) = O(a^{3/2}), \quad 
\phi(x), \bar{\phi}(x) = O(a), \quad \vec{H}(x) = O(a^2), \nn \\
 & & Q=O(a^{1/2}). 
\label{order_of_a}
\eea

The action (\ref{4DN2_Sn1}) takes the almost same form as the one presented in the paper \cite{sugino} 
except the terms containing $\Delta\vec{\Phi}(x)$. 
The result of the $Q$-action in (\ref{4DN2_Sn1}) becomes 
\bea
S^{{\rm LAT}}_{4D{\cal N}=2} & = & \frac{1}{2g_0^2}\sum_x \, \tr\left[
\frac14 [\phi(x), \,\bar{\phi}(x)]^2 + \vec{H}(x)\cdot\vec{H}(x) 
-i\vec{H}(x)\cdot(\vec{\Phi}(x)+\Delta\vec{\Phi}(x)) \right. \nn \\
 & & 
+\sum_{\mu=1}^4\left(\phi(x)-U_{\mu}(x)\phi(x+\hat{\mu})U_{\mu}(x)^{\dagger}
\right)\left(\bar{\phi}(x)-U_{\mu}(x)\bar{\phi}(x+\hat{\mu})
U_{\mu}(x)^{\dagger}\right) \nn \\
 & &  -\frac14 \eta(x)[\phi(x), \,\eta(x)] 
- \vec{\chi}(x)\cdot [\phi(x), \,\vec{\chi}(x)] \nn \\
 & & 
-\sum_{\mu=1}^4\psi_{\mu}(x)\psi_{\mu}(x)\left(\bar{\phi}(x)  + 
U_{\mu}(x)\bar{\phi}(x+\hat{\mu})U_{\mu}(x)^{\dagger}\right) 
+ i\vec{\chi}(x)\cdot Q(\vec{\Phi}(x)  +\Delta\vec{\Phi}(x)) \nn \\
 & & \left. \frac{}{} -i\sum_{\mu=1}^4\psi_{\mu}(x)\left(\eta(x)-
U_{\mu}(x)\eta(x+\hat{\mu})U_{\mu}(x)^{\dagger}\right)\right]. 
\label{4DN2_Sn2}
\eea
Note that since $\vec{H}$ is traceless, the classical vacua are determined by 
\beq
\vec{\Phi}(x) + \Delta\vec{\Phi}(x) - \left(\frac{1}{N}\tr[\vec{\Phi}(x) + \Delta\vec{\Phi}(x)]\right){\bf 1}_N =0. 
\label{vacuaSU(N)}
\eeq 
As seen in appendix A, however its solutions are 
not unique and distribute continuously from $U_{\mu\nu}(x)=1$. 
Thus, a manipulation using the admissibility conditions would not work to single out the minimum 
$U_{\mu\nu}(x) =1$ as in \cite{sugino2}. 

On the other hand, the equation  
\beq
\Phi(x) + \Delta\Phi(x)=0  
\label{vacuaU(N)}
\eeq 
for $G=\mbox{SU}(N)$ has the unique solution $U_{\mu\nu}(x)=1$ 
with appropriately chosen $r$ 
%
in the range\footnote{In appendix B, 
we prove the unique solution $U_{\mu\nu}(x)=1$ for the equations $\vec{\Phi}(x)+\Delta\vec{\Phi}(x)=0$ 
in the case limited to (\ref{parameter_r}). 
A priori, the argument itself does not exclude the possibility of the unique solution 
for the parameter $r$ outside of the range (\ref{parameter_r}).} 
\beq
r=\cot \varphi: 
\quad 0<\varphi\leq \frac{\pi}{2N},
\label{parameter_r}
\eeq 
as shown in appendix B. 

Here, we extend $\vec{\chi}(x)$, $\vec{H}(x)$ to the hermitian matrices $\vec{\hat{\chi}}(x)$ , $\vec{\hat{H}}(x)$ 
with nonzero trace parts to 
introduce the variables $\vec{\chi}^{(0)}(x)$, $\vec{H}^{(0)}(x)$ 
proportional to the unit matrix:
\bea
 & & \vec{\chi}^{(0)}(x) = \vec{\underline{\chi}}^{(0)}(x) \, {\bf 1}_N, \qquad 
     \vec{H}^{(0)}(x) = \vec{\underline{H}}^{(0)}(x) \, {\bf 1}_N \nn \\
 & & \vec{\hat{H}}(x) = \vec{H}(x) + \vec{H}^{(0)}(x), \qquad \vec{\hat{\chi}}(x) = \vec{\chi}(x) + \vec{\chi}^{(0)}(x). 
\eea
The fields with the uniderline mean the coefficients proportional to the unit matrix. 
The $Q$-transformation (\ref{Q_lattice}) of $\vec{\chi}(x)$ and $\vec{H}(x)$ is naturally extended to 
\beq
Q\vec{\hat{\chi}}(x)= \vec{\hat{H}}(x), \qquad Q\vec{\hat{H}}(x) = [\phi(x), \vec{\hat{\chi}}(x)] 
\eeq
with 
\beq
Q\vec{\chi}^{(0)}(x) = \vec{H}^{(0)}(x), \qquad Q\vec{H}^{(0)}(x) = 0. 
\eeq

The lattice action is modified as 
\bea
\hat{S}^{{\rm LAT}}_{4D{\cal N}=2} & = & Q\frac{1}{2g_0^2}\sum_x \, \tr\left[ 
\frac14 \eta(x)\, [\phi(x), \,\bar{\phi}(x)] 
-i\vec{\hat{\chi}}(x)\cdot\left(\vec{\Phi}(x) + \Delta\vec{\Phi}(x)\right)
+\vec{\hat{\chi}}(x)\cdot\vec{\hat{H}}(x)\right. \nn \\
 & & \hspace{2cm}\left. \frac{}{} 
+i\sum_{\mu=1}^2\psi_{\mu}(x)\left(\bar{\phi}(x) - 
U_{\mu}(x)\bar{\phi}(x+\hat{\mu})U_{\mu}(x)^{\dagger}\right)\right] 
\label{U(N)lat_4DN=2_S}   \\
& = & \frac{1}{2g_0^2}\sum_x \, \tr\left[
\frac14 [\phi(x), \,\bar{\phi}(x)]^2 + \vec{\hat{H}}(x)^2 
-i\vec{\hat{H}}(x)\cdot\left(\vec{\Phi}(x) + \Delta \vec{\Phi}(x)\right) \right. \nn \\
 & & \hspace{1.5cm}
+\sum_{\mu=1}^2\left(\phi(x)-U_{\mu}(x)\phi(x+\hat{\mu})U_{\mu}(x)^{\dagger}
\right)\left(\bar{\phi}(x)-U_{\mu}(x)\bar{\phi}(x+\hat{\mu})
U_{\mu}(x)^{\dagger}\right) \nn \\
 & & \hspace{1.5cm} -\frac14 \eta(x)[\phi(x), \,\eta(x)] 
- \vec{\chi}(x)\cdot [\phi(x), \,\vec{\chi}(x)] \nn \\
 & & \hspace{1.5cm}
-\sum_{\mu=1}^2\psi_{\mu}(x)\psi_{\mu}(x)\left(\bar{\phi}(x)  + 
U_{\mu}(x)\bar{\phi}(x+\hat{\mu})U_{\mu}(x)^{\dagger}\right) \nn \\
 & & \hspace{1.5cm}   + i\vec{\hat{\chi}}(x) \cdot Q\left(\vec{\Phi}(x) + \Delta\vec{\Phi}(x)\right)\nn \\
 & & \hspace{1.5cm}\left. \frac{}{}
-i\sum_{\mu=1}^2\psi_{\mu}(x)\left(\eta(x)-
U_{\mu}(x)\eta(x+\hat{\mu})U_{\mu}(x)^{\dagger}\right)\right]. 
\label{U(N)lat_N=2_S2}
\eea

Due to the trace part of $\vec{\hat{H}}(x)$, the minimum of the gauge part is uniquely determined by 
(\ref{vacuaU(N)}) with $r = \cot \varphi$ satisfying (\ref{parameter_r}).
On the other hand, the kinetic term of $\vec{\hat{\chi}}(x)$ becomes 
\beq
\frac{1}{2g_0^2}\sum_x \left\{\tr \left[i\vec{\chi}(x)\cdot Q(\vec{\Phi}(x) + \Delta\vec{\Phi}(x))\right] 
+ i\vec{\underline{\chi}}^{(0)}(x)\cdot Q\,\tr(\vec{\Phi}(x) + \Delta\vec{\Phi}(x))\right\}.   
\eeq
Since the second term in the brace is of the order $O(a^6)$, 
it vanishes in the contiuum limit and $\vec{\underline{\chi}}^{(0)}(x)$ become fermion zero-modes. 
If we integrate out $\vec{\underline{\chi}}^{(0)}(x)$, we will have the nontrivial constraints 
\beq
0 = Q\, \tr(\vec{\Phi}(x) + \Delta\vec{\Phi}(x))
\eeq
leading 
\bea
0 & = &  \tr \left[F_{14}(x)(D_1\psi_4(x)-D_4\psi_1(x)) + F_{23}(x)(D_2\psi_3(x)-D_3\psi_2(x))\right], \nn \\
0 & = &  \tr \left[F_{24}(x)(D_2\psi_4(x)-D_4\psi_2(x)) + F_{31}(x)(D_3\psi_1(x)-D_1\psi_3(x))\right], \nn \\
0 & = &  \tr \left[F_{34}(x)(D_3\psi_4(x)-D_4\psi_3(x)) + F_{12}(x)(D_1\psi_2(x)-D_2\psi_1(x))\right]
\eea
at the nontrivial leading order $O(a^{9/2})$ in the continuum. 
Because such constraints are not imposed in the target continuum theory, we should avoid obtaining them. 
In order to do so, we soak up the would-be fermion zero-modes in the path-integral to consider the 
measure 
\beq
\dd \mu_{{\cal N}=2} \equiv \dd \mu_{{\rm SU}(N)\, {\cal N}=2}
 \prod_{x, {\bA}}\left(\dd \underline{H}_{\bA}^{(0)}(x)\dd \underline{\chi}_{\bA}^{(0)}(x)\, 
\underline{\chi}_{\bA}^{(0)}(x)\right)  
\eeq
with $\dd \mu_{{\rm SU}(N)\, {\cal N}=2}$ being the measure for the SU$(N)$ variables. 
Note that $\dd \underline{\chi}_{\bA}^{(0)}(x)\, \underline{\chi}_{\bA}^{(0)}(x)$ is U$(1)_R$ invariant, because 
$\dd \underline{\chi}_{\bA}^{(0)}(x)$ transforms same as the derivative 
$\partial /\partial \underline{\chi}_{\bA}^{(0)}(x)$. 
However, the insertion of the would-be zero-modes violates the $Q$ invariance as 
\beq
Q\left(\dd \underline{H}_{\bA}^{(0)}(x)\dd \underline{\chi}_{\bA}^{(0)}(x)\, \underline{\chi}_{\bA}^{(0)}(x)\right)
= -\dd \underline{H}_{\bA}^{(0)}(x)\dd \underline{\chi}_{\bA}^{(0)}(x)\,\underline{H}_{\bA}^{(0)}(x), 
\label{Q_measure}
\eeq
although the action (\ref{U(N)lat_4DN=2_S}) is manifestly $Q$ invariant. 

Here we consider the observables consisting the operators in the SU$(N)$ sector i.e. independent of 
$\vec{H}^{(0)}(x)$ and $\vec{\chi}^{(0)}(x)$. Let us write the action as 
\bea
\hat{S}^{{\rm LAT}}_{4D{\cal N}=2} & = & S^{{\rm LAT}}_{4D{\cal N}=2} + 
      \frac{N}{2g_0^2}\sum_x\left[\vec{\underline{H}}^{(0)}(x)^2 
           -i\vec{\underline{H}}^{(0)}(x)\cdot \frac{1}{N}\tr(\vec{\Phi}(x) + \Delta\vec{\Phi}(x)) \right. \nn \\
            & & \hspace{3cm}\left. 
                 +i\vec{\underline{\chi}}^{(0)}(x)\cdot Q\frac{1}{N}\tr (\vec{\Phi}(x) + \Delta\vec{\Phi}(x))\right], 
\label{action_bunri}
\eea
so that the dependence of $\vec{\underline{H}}^{(0)}(x)$ and $\vec{\underline{\chi}}^{(0)}(x)$ can be explicitly seen. 
{}From (\ref{Q_measure}), (\ref{action_bunri}), 
the $Q$-transformation of $\dd \mu_{{\cal N}=2}\,e^{-\hat{S}^{{\rm LAT}}_{4D{\cal N}=2}}$ leads 
\bea
 & & \int Q\left(\dd \mu_{{\cal N}=2}\,e^{-\hat{S}^{{\rm LAT}}_{4D{\cal N}=2}}\right) = 
  \int \dd \mu_{{\rm SU}(N)\, {\cal N}=2}\, e^{-S^{{\rm LAT}}_{{\cal N}=2}}\, 
   \left(\prod_{x, {\bA}}\dd \underline{H}_{\bA}^{(0)}(x)\right) \nn \\
  &  & \hspace{2cm}\times \sum_x \left[\frac{i}{2g_0^2} \vec{\underline{H}}^{(0)}(x)\cdot Q\,
                           \tr(\vec{\Phi}(x) + \Delta\vec{\Phi}(x))\right] \nn \\
 & & \hspace{2cm}\times \exp\left\{-\frac{N}{2g_0^2}\sum_x \left[ \vec{\underline{H}}^{(0)}(x)^2 
            -i \vec{\underline{H}}^{(0)}(x)\cdot \frac{1}{N}\tr(\vec{\Phi}(x) + \Delta\vec{\Phi}(x))\right]\right\}
\eea
after integrating out $\vec{\underline{\chi}}^{(0)}$. As the result of the integration of $\vec{\underline{H}}^{(0)}$, we obtain 
\bea
\int Q\left(\dd \mu_{{\cal N}=2}\,e^{-\hat{S}^{{\rm LAT}}_{4D{\cal N}=2}}\right) & = & 
  \int \dd \mu_{{\rm SU}(N)\, {\cal N}=2}\, e^{-S^{{\rm LAT}}_{{\cal N}=2}}\, 
   e^{-\frac{1}{8Ng_0^2}\sum_x\left[\tr(\vec{\Phi}(x) + \Delta\vec{\Phi}(x))\right]^2} \nn \\
 & & \times Q\sum_x\frac{-1}{8Ng_0^2}\left[\tr(\vec{\Phi}(x) + \Delta\vec{\Phi}(x))\right]^2,  
\eea
which means that the insertion of the would-be fermion zero-modes is equivalent to adding the supersymmetry breaking term 
\beq
\Delta S = \frac{1}{8Ng_0^2}\sum_x\left[\tr(\vec{\Phi}(x) + \Delta\vec{\Phi}(x))\right]^2
\label{Delta_S}
\eeq
to the $Q$ invariant action $S^{{\rm LAT}}_{4D{\cal N}=2}$ (\ref{4DN2_Sn1}). 
$\Delta S$ supplies the trace part of $\vec{\Phi}(x) + \Delta\vec{\Phi}(x)$ leading the condition for the minima (\ref{vacuaU(N)}) 
to resolve the degeneracy. In the continuum limit, $\Delta S$ becomes irrelevant, because 
$\tr(\vec{\Phi}(x) + \Delta\vec{\Phi}(x))$ is of the order $O(a^4)$.


\paragraph{Surplus Modes}
The action (\ref{U(N)lat_4DN=2_S}), equivalently to $S^{{\rm LAT}}_{4D{\cal N}=2}+ \Delta S$, 
resolves the vacuum degeneracy to fix the single vacuum $U_{\mu\nu}(x)=1$. 
Because $U_{\mu\nu}(x)=1$ is equivalent to $U_{\mu}(x)=1$ modulo gauge transformations 
on the infinite lattice $x\in {\bf Z}^4$, 
the expansion 
\beq
U_{\mu}(x) = 1 + iaA_{\mu}(x) +\cdots 
\label{expansion_U} 
\eeq
is justified to treat fluctuations around the minimum.   

After the weak field expansion (\ref{expansion_U}), 
the fermion kinetic terms are expressed as 
\bea
S_f^{(2)} & = & \frac{a^4}{2g_0^2}\sum_{x, \mu}\tr\left[
-\frac12\Psi(x)^T (P_{\mu}+\gamma_{\mu})\Delta_{\mu}\Psi(x) 
+\frac12\Psi(x)^T (P_{\mu}-\gamma_{\mu})\Delta^*_{\mu}\Psi(x)\right] \nn \\
 & = & \frac{a^4}{2g_0^2}\sum_{x, \mu}\tr\left[
-\frac12\Psi(x)^T\gamma_{\mu}(\Delta_{\mu}+\Delta^*_{\mu})\Psi(x) 
-a\frac12\Psi(x)^TP_{\mu}\Delta_{\mu}\Delta^*_{\mu}\Psi(x)\right], 
\label{wilson_like_N=2}
\eea
where the fermion fields were rescaled by $a^{3/2}$ as indicated in  
(\ref{order_of_a}) and combined as 
\beq
\Psi^T = \left( \psi_1, \cdots, \psi_4, \chi_1, \chi_2, \chi_3, 
\frac12\eta \right). 
\label{Psi_N=2}
\eeq
The transpose operation acts only on the spinor indices. 
$\Delta_{\mu}$ and $\Delta^*_{\mu}$ represent forward and backward difference  
operators respectively: 
\beq
\Delta_{\mu}f(x) \equiv \frac{1}{a}\left(f(x+\hat{\mu})-f(x)\right), \quad 
\Delta^*_{\mu}f(x) \equiv \frac{1}{a}\left(f(x)-f(x-\hat{\mu})\right).
\eeq
The explicit form of the matrices $\gamma_{\mu}$, $P_{\mu}$ is given in appendix C.1.  
They satisfy 
\beq
\{ \gamma_{\mu}, \gamma_{\nu} \} = -2\delta_{\mu\nu}, \quad 
\{ P_{\mu}, P_{\nu}\} = 2\delta_{\mu\nu}, \quad 
\{ \gamma_{\mu}, P_{\mu} \} = 0, 
\eeq
while $\gamma_{\mu}$ and $P_{\nu}$ with $\mu\neq\nu$ do not anticommute: 
\beq
\{ \gamma_{\mu}, P_{\nu} \} \neq 0 \quad \mbox{for  }\mu \neq \nu. 
\label{gamma_P_diff}
\eeq
Notice its difference from the two-dimensional ${\cal N}=2, 4$ cases 
in the paper \cite{sugino2}, where 
\beq
\{ \gamma_{\mu}, P_{\nu}\}=0 \quad \mbox{for} \quad  
\forall \mu, \nu =1, 2 
\label{no_surplus}
\eeq
leads no fermion doublers.  

The kernel of the kinetic term (\ref{wilson_like_N=2}) is written in the momentum space 
$-\pi /a\leq q_{\mu}<\pi/a$ as 
\beq
\tilde{D}(q)=\sum_{\mu=1}^4
\left[-i\gamma_{\mu}\frac{1}{a}\sin \left(q_{\mu}a\right)
+2P_{\mu}\frac{1}{a}\sin^2 \left(\frac{q_{\mu}a}{2}\right)\right]. 
\label{Dirac_op}
\eeq
Due to the property (\ref{gamma_P_diff}), 
it is not clear if $\tilde{D}(q)$ has zero-modes only at the origin 
$(q_1, \cdots, q_4)=(0,\cdots, 0)$. Indeed, $\tilde{D}(q)$ has zero-eigenvalues at the points 
\bea
 & & (q_1, q_2, q_3, q_4)  = (0,0,0,0), \quad (\pm\frac{\pi}{2a}, \mp\frac{\pi}{2a}, 0,0), 
\quad (\pm\frac{\pi}{2a}, 0,\mp\frac{\pi}{2a},0), 
\quad (\pm\frac{\pi}{2a}, 0,0,\mp\frac{\pi}{2a}), \nn \\
 &  & \quad (0,\pm\frac{\pi}{2a}, \mp\frac{\pi}{2a},0), 
\quad (0,\pm\frac{\pi}{2a}, 0,\mp\frac{\pi}{2a}), 
\quad (0,0,\pm\frac{\pi}{2a}, \mp\frac{\pi}{2a}), \nn \\
 & & \quad (\pm\frac{\pi}{2a}, \mp\frac{\pi}{2a},\pm\frac{\pi}{2a}, \mp\frac{\pi}{2a}), 
\quad (\pm\frac{\pi}{2a}, \pm\frac{\pi}{2a},\mp\frac{\pi}{2a},\mp\frac{\pi}{2a}), 
\quad (\pm\frac{\pi}{2a}, \mp\frac{\pi}{2a},\mp\frac{\pi}{2a},\pm\frac{\pi}{2a}), \nn \\
 & & \quad (\pm\frac{\pi}{3a}, \pm\frac{\pi}{3a},\pm\frac{\pi}{3a},\mp\frac{2\pi}{3a}), 
\quad (\pm\frac{\pi}{3a}, \pm\frac{\pi}{3a},\mp\frac{2\pi}{3a},\pm\frac{\pi}{3a}),
\quad (\pm\frac{\pi}{3a}, \mp\frac{2\pi}{3a},\pm\frac{\pi}{3a},\pm\frac{\pi}{3a}), \nn \\
 & & \quad (\mp\frac{2\pi}{3a}, \pm\frac{\pi}{3a},\pm\frac{\pi}{3a},\pm\frac{\pi}{3a}). 
\label{naive_surplus}
\eea
The modes other than $(q_1, \cdots, q_4)=(0,\cdots, 0)$ are surplus, and 
we do not want them in order to get the desired continuum theory.  
Since the surplus modes are not exact zero-modes of the full action 
({\em they are only of the quadratic term}), 
for each of them a corresponding supersymmetric partner exists in the bosonic sector 
in consequence of the exact supersymmetry $Q$. 
(In the weak field expansion (\ref{expansion_U}), the effect of $\Delta S$ is irrelevant.)
 
Let us see the bosonic surplus modes. 
We may consider quadratic terms of $A_{\mu}(x)$ in 
$\tr \left[\vec{\Phi}(x)+\Delta\vec{\Phi}(x)\right]^2$ 
after the expansion (\ref{expansion_U}), i.e. linear terms of $A_{\mu}(x)$ in $\vec{\Phi}(x)$. 
Note that terms from $\Delta\vec{\Phi}(x)$ 
yield irrelevant interaction terms instead of the gauge kinetic term, and we can neglect them 
in the analysis here. 
In terms of the Fourier transformed basis  
\beq
\Phi_{\bA}(x) =\int^{\pi/a}_{-\pi/a}\frac{\dd^4q}{(2\pi)^4}\, e^{iaq\cdot x}\, \tilde{\Phi}_{\bA}(q), 
\eeq
$\vec{\Phi}$ becomes 
\bea
\tilde{\Phi}_1(q) & = & 2a \left[(e^{iaq_1}-1)\tilde{A}_4(q) -(e^{iaq_4}-1)\tilde{A}_1(q) 
+(e^{iaq_2}-1)\tilde{A}_3(q) -(e^{iaq_3}-1)\tilde{A}_2(q)\right] \nn \\
 & & +O(\tilde{A}^2) \nn \\
\tilde{\Phi}_2(q) & = & 2a \left[(e^{iaq_2}-1)\tilde{A}_4(q) -(e^{iaq_4}-1)\tilde{A}_2(q) 
+(e^{iaq_3}-1)\tilde{A}_1(q) -(e^{iaq_1}-1)\tilde{A}_3(q)\right] \nn \\
 & & +O(\tilde{A}^2) \nn \\
\tilde{\Phi}_3(q) & = & 2a \left[(e^{iaq_3}-1)\tilde{A}_4(q) -(e^{iaq_4}-1)\tilde{A}_3(q) 
+(e^{iaq_1}-1)\tilde{A}_2(q) -(e^{iaq_2}-1)\tilde{A}_1(q)\right] \nn \\
 & & +O(\tilde{A}^2). 
\eea
We shall see that there appear at the momenta (\ref{naive_surplus}) 
the zero-modes $\tilde{A}_\mu(q)$ satisfying 
\beq
\vec{\tilde{\Phi}}(q)=0 + O(\tilde{A}^2). 
\label{Phi_zero}
\eeq
The case $q=0$ is trivial. For the rest, 
as an example, let us consider 
the case $q_{(12)}\equiv (\frac{\pi}{2a}, -\frac{\pi}{2a},0,0)$. 
The solution of 
(\ref{Phi_zero}) is 
\beq
\left(\tilde{A}_1^{\alpha}(q_{(12)}), \tilde{A}_2^{\alpha}(q_{(12)}), \tilde{A}_3^{\alpha}(q_{(12)}), 
\tilde{A}_4^{\alpha}(q_{(12)})\right) = (K^{\alpha}, iK^{\alpha}, L^{\alpha}, -iL^{\alpha} )
\eeq
with $\alpha=1, \cdots, N^2-1$ being the index of a basis of the gauge generators 
and $K^{\alpha}, L^{\alpha}$ complex numbers. 
Because $K^{\alpha}$ is absorbed by the linearized gauge transformation 
\beq
\delta \tilde{A}_{\mu}^{\alpha}(q_{(12)})=
(e^{iaq_{(12)\mu}}-1) \frac{1}{1-i} K^{\alpha}, 
\eeq
only one degree of freedom $L^{\alpha}$ is dynamical. 
Correspondingly, the Dirac operator (\ref{Dirac_op}) at $q=q_{(12)}$ has the two eigenvectors for 
zero-eigenvalue: 
\beq
(0,0,1, -i, 0,0,0,0)^T, \quad (0,0,0,0,1,-i,0,0)^T \nn 
\eeq
meaning one on-shell degree of freedom for fermions. 
Hence, bosonic and fermionic degrees of freedom for the surplus mode $q=q_{(12)}$ are 
balanced\footnote{The on-shell degrees of freedom of the mode $q=q_{(12)}$ are smaller 
compared with the zero-mode $q=0$ having two on-shell degrees of freedom. 
Thus, we consider the mode $q=q_{(12)}$ not identical with the so-called doublers, and use the term 
``surplus modes". 
If it was the doubler, the Dirac operator $\tilde{D}(q_{(12)})$ would become the zero matrix. 
However, it is not the case. $\tilde{D}(q_{(12)})$ has two zero-eigenvalues {\it as well as 
six nonzero-eigenvalues}. The situation is same for the other surplus modes. 
The notion of the surplus modes is broader than that of the doublers. 
Nonexistence of the surplus modes leads nonexistence of the doublers, but the reverse does not hold.}. 
The analysis for the other modes is similarly given.   

The appearence of the surplus modes seems related to exact realization of a lattice counterpart of 
the topological term 
$\int \dd^4 x\, \varepsilon_{\mu\nu\lambda\rho}\, \tr F_{\mu\nu}(x)F_{\lambda\rho}(x)$. 
Writing as 
\bea
 & & \Phi_1(x) +\Delta\Phi_1(x) = 
{\cal F}_{1 4}(x) + {\cal F}_{23}(x), \nn \\
 & & \Phi_2(x) +\Delta\Phi_2(x) = 
{\cal F}_{2 4}(x) + {\cal F}_{31}(x), \nn \\
 & & \Phi_3(x) +\Delta\Phi_3(x) = 
{\cal F}_{3 4}(x) + {\cal F}_{12}(x), \nn \\
 & & {\cal F}_{\mu\nu}(x)  \equiv  -i(U_{\mu\nu}(x)-U_{\nu\mu}(x)) -rW_{\mu\nu}(x), 
\eea
${\cal F}_{\mu\nu}(x)$ reduce the continuum field strengths $F_{\mu\nu}(x)$ in the limit $a\limit 0$. 
It is easy to see that the surplus modes do not arise from $\tr {\cal F}_{\mu\nu}(x)^2$ 
in the weak field expansion (\ref{expansion_U}) 
of $\tr\left[\vec{\Phi}(x)+\Delta\vec{\Phi}(x)\right]^2$, 
which suggests that 
the cross term 
\beq
\tr{\cal F}_{14}(x){\cal F}_{23}(x) + \tr{\cal F}_{24}(x){\cal F}_{31}(x) + 
\tr{\cal F}_{34}(x){\cal F}_{12}(x)
\eeq
contains extra dynamical degrees of freedom corresponding to the surplus modes. 
Thus it is not  
exactly topological at the lattice level in the sense that an arbitrary small variation of 
the quantity does not always vanish. 
For the case of the gauge group U(1), the topological term has been rigorously constructed 
on the lattice in refs. \cite{luscher}. 
Extension of the construction to the nonabelian case would be quite helpful to resolve our problem. 

Due to the supersymmetric property, it might seem that bosonic and fermionic contributions 
to the surplus modes cancel each other among loop diagrams in the perturbation theory, 
and thus they might be harmless as long as considering observables 
which do not allow the surplus modes to propagate from the corresponding external lines. 
In order to clarify it, however we need an explicit analysis of the lattice perturbation theory. 
Because it is not a main issue of this paper, we will examine the above speculation elsewhere.  
  
Note that our model realizes the U$(1)_R$ symmetry (\ref{U1R}) instead 
of the fermion number conservation at the lattice level. 
Anomaly will be induced with respect to the fermion number conservation 
that is broken by the lattice artifact. 
The situation is opposite to the conventional treatment of the four-dimensional 
${\cal N}=2$ SYM theory (For example, see \cite{seiberg-witten}.), 
where U$(1)_R$ becomes anomalous and the fermion number 
symmetry U$(1)_J$ is preserved at the quantum level. 
In order to remedy it and to reproduce the U$(1)_R$ anomaly correctly, 
we will have to introduce suitable counter terms to the lattice action 
to cancel the fermion number anomaly.     
In this paper, we do not pursue the four-dimensional model itself further. 
Instead, it is utilized as an intermediate step 
to construct lower-dimensional models {\it not containing 
surplus modes}, after making a slight modification to the four-dimensional model 
in the next subsection. 
This is a main issue of this paper. 

\subsection{Slightly Modified Model for 4D ${\cal N}=2$}

We slightly modify the action $S^{{\rm LAT}}_{4D{\cal N}=2}+ \Delta S$ with the replacement 
$\vec{\Phi} \limit \vec{\Phi}'$, $\Delta\vec{\Phi} \limit \Delta\vec{\Phi}'$, where 
\bea
\Phi'_1(x) & \equiv & -i[U_{4, -1}(x) - U_{-1, 4}(x) + U_{23}(x) - U_{32}(x)], \nn \\
\Phi'_2(x) & \equiv & -i[U_{4, -2}(x) - U_{-2, 4}(x) + U_{31}(x) - U_{13}(x)], \nn \\ 
\Phi'_3(x) & \equiv & -i[U_{4, -3}(x) - U_{-3, 4}(x) + U_{12}(x) - U_{21}(x)], \nn \\
\Delta\Phi'_1(x) & \equiv & -r[W_{4, -1}(x) + W_{23}(x)], \nn \\
\Delta\Phi'_2(x) & \equiv & -r[W_{4, -2}(x) + W_{31}(x)], \nn \\
\Delta\Phi'_3(x) & \equiv & -r[W_{4, -3}(x) + W_{12}(x)]. 
\label{Phip}
\eea
The parameter $r$ is chosen in the range (\ref{parameter_r}) again. 
The link variables along the $-\mu$ direction are defined as 
\beq
U_{-\mu}(x) \equiv U_{\mu}(x-\hat{\mu})^{\dagger}, 
\eeq
and then plaquettes $U_{\mu\nu}(x)$ for $\mu, \nu=\pm 1, \cdots, \pm 4$ are expressed as 
\beq
U_{\mu\nu}(x) = U_{\mu}(x)U_{\nu}(x+\hat{\mu})U_{-\mu}(x+\hat{\mu}+\hat{\nu})
U_{-\nu}(x+\hat{\nu}) \qquad (\mu, \nu =\pm 1, \cdots, \pm 4). 
\eeq

The fermion kinetic term of the modified model takes the form 
\bea
S^{(2)}_f & = & \frac{a^4}{2g_0^2}\sum_{x}\tr\left[\sum_{\mu =1}^4
\left\{-\frac12\Psi(x)^T\gamma_{\mu}(\Delta_{\mu}+\Delta^*_{\mu})\Psi(x) 
-a\frac12\Psi(x)^TP'_{\mu}\Delta_{\mu}\Delta^*_{\mu}\Psi(x) \right\}\right. \nn \\
  & & \left. \frac{}{} \hspace{1.8cm}
+ia\sum_{\bA =1}^3(\chi_{\bA}(x)\Delta_4\Delta^*_{\bA}\psi_{\bA}(x) 
-\psi_{\bA}(x)\Delta_{\bA}\Delta^*_4\chi_{\bA}(x) )
\right],   
\label{Sf_modified}
\eea
with the matrices $P'_{\mu}$, whose explicit form is given in appendix C.2, satisfying 
\bea
 & & \{ \gamma_\mu, \gamma_\nu \}=-2\delta_{\mu\nu}, \quad 
     \{ P'_i, P'_j\} = 2\delta_{ij}, \quad \{ \gamma_i, P'_j \} =0, \nn \\
 & & \{ P'_4, P'_i \} \neq 0, \quad \{ \gamma_4, P'_i \} \neq 0, \quad \{ \gamma_i, P'_4 \} \neq 0, 
\quad (P'_4)^2 =1 \quad (i, j=1,2,3). 
\label{gamma_P_algebra}
\eea
Note that $\gamma_i$ anticommute with $P'_j$ ($i, j= 1, 2, 3$). 
{}From the same reason as (\ref{no_surplus}) in the case \cite{sugino2}, the fermion kinetic term 
(\ref{Sf_modified}) has the unique zero-mode at $q_1=q_2=q_3=0$, 
after the dimensional reduction with respect to the fourth 
direction $x_4$ which kills the terms containing $\Delta_4$ or $\Delta^*_4$.    

The four-dimensional modified action still has the surplus modes. However, for each of them 
the momentum has the nonzero 
fourth component ($q_4\neq 0$), and they are removed in the models obtained by the 
dimensional reduction with respect to the fourth direction. 
In this way, 
we can define ${\cal N}=4$ lattice SYM models in two and three dimensions {\it without 
surplus modes}
.
We will see the details about the dimensionally reduced lattice models in the next section.

\setcounter{equation}{0}
\section{Dimensionally Reduced ${\cal N}=4$ Models in Two and Three Dimensions} 
\label{sec:23DN4}
We discuss on two- and three-dimensional ${\cal N}=4$ SYM models 
obtained by dimensional reduction from the slightly 
modified ${\cal N}=2$ model in four dimensions.  
  
\subsection{3D ${\cal N}=4$ Case}
Here, we consider the dimensional reduction with respect to $x_4$ in the slightly 
modified model in section 2.2. The result leads ${\cal N}=4$ SYM on the three-dimensional lattice, 
whose action $S^{{\rm LAT}}_{3D{\cal N}=4} + \Delta S'$ is expressed as  
\bea
S^{{\rm LAT}}_{3D{\cal N}=4} & = & Q\frac{1}{2g_0^2}\sum_x \, \tr\left[ 
\frac14 \eta(x)\, [\phi(x), \,\bar{\phi}(x)] -i\vec{\chi}(x)\cdot(\vec{\Phi}'(x) 
+ \Delta\vec{\Phi}'(x))
+\vec{\chi}(x)\cdot\vec{H}(x)\right. \nn \\
 & & \hspace{2cm}
+i\sum_{\mu=1}^3\psi_{\mu}(x)\left(\bar{\phi}(x) - 
U_{\mu}(x)\bar{\phi}(x+\hat{\mu})U_{\mu}(x)^{\dagger}\right)\nn \\
 & & \hspace{2cm} \left.\frac{}{}+i\psi_{4}(x)\left(\bar{\phi}(x) - 
V(x)\bar{\phi}(x)V(x)^{\dagger}\right)\right], 
\label{3DN4_S1}
\eea
where $x$ represents the three-dimensional lattice site $(x_1, x_2, x_3)$, and 
$V(x)\equiv U_4(x)$, $V(x)^{\dagger}=U_{-4}(x)$ become sitting at the site $x$.
$\Delta S'$ is given by the formula (\ref{Delta_S}) with the replacement $\vec{\Phi}(x) \limit \vec{\Phi}'(x)$, 
$\Delta \vec{\Phi}(x) \limit \Delta \vec{\Phi}'(x)$: 
\beq
\Delta S' = \frac{1}{8Ng_0^2}\sum_x\left[\tr(\vec{\Phi}'(x) + \Delta\vec{\Phi}'(x))\right]^2.
\label{Delta_S'}
\eeq
Also, plaquettes $U_{-\mu, 4}(x)$ collapse to the links: 
\beq
U_{-\mu, 4}(x) = U_{\mu}(x-\hat{\mu})^{\dagger} V(x-\hat{\mu})U_{\mu}(x-\hat{\mu})V(x)^{\dagger}
=U_{4, -\mu}(x)^{\dagger} \qquad (\mu =1, 2, 3).  
\label{rule_dimred}
\eeq
The manifest supersymmetry is given by the dimensional reduction of (\ref{Q_lattice}). 
Namely, for the cases $\mu = 1, 2, 3$ of (\ref{Q_lattice}), the form does not change, 
while for the $\mu =4$ case it becomes 
\beq
QV(x)=i\psi_4(x)V(x), \quad Q\psi_4(x)=i\psi_4(x)\psi_4(x)-i\left(\phi(x)-V(x)\phi(x)V(x)^\dagger\right). 
\eeq 

The action $S^{{\rm LAT}}_{3D{\cal N}=4}$ is explicitly written as 
\bea
S^{{\rm LAT}}_{3D{\cal N}=4} & = & \frac{1}{2g_0^2}\sum_x \, \tr\left[
\frac14 [\phi(x), \,\bar{\phi}(x)]^2 + \vec{H}(x)\cdot\vec{H}(x) 
-i\vec{H}(x)\cdot(\vec{\Phi}'(x)+\Delta\vec{\Phi}'(x)) \right. \nn \\
 & & 
+\sum_{\mu=1}^3\left(\phi(x)-U_{\mu}(x)\phi(x+\hat{\mu})U_{\mu}(x)^{\dagger}
\right)\left(\bar{\phi}(x)-U_{\mu}(x)\bar{\phi}(x+\hat{\mu})
U_{\mu}(x)^{\dagger}\right) \nn \\
 & & + \left(\phi(x)-V(x)\phi(x)V(x)^{\dagger}\right)
\left(\bar{\phi}(x)-V(x)\bar{\phi}(x)V(x)^{\dagger}\right) \nn \\
 & &  -\frac14 \eta(x)[\phi(x), \,\eta(x)] 
- \vec{\chi}(x)\cdot [\phi(x), \,\vec{\chi}(x)] \nn \\
 & & 
-\sum_{\mu=1}^3\psi_{\mu}(x)\psi_{\mu}(x)\left(\bar{\phi}(x)  + 
U_{\mu}(x)\bar{\phi}(x+\hat{\mu})U_{\mu}(x)^{\dagger}\right) \nn \\
 & & 
-\psi_4(x)\psi_4(x)\left(\bar{\phi}(x) + V(x)\bar{\phi}(x)V(x)^{\dagger}\right)
+ i\vec{\chi}(x)\cdot Q(\vec{\Phi}'(x)  +\Delta\vec{\Phi}'(x)) \nn \\
 & & -i\sum_{\mu=1}^3\psi_{\mu}(x)\left(\eta(x)-
U_{\mu}(x)\eta(x+\hat{\mu})U_{\mu}(x)^{\dagger}\right) \nn \\ 
 & & \left. \frac{}{} -i\psi_4(x)\left(\eta(x)-V(x)\eta(x)V(x)^{\dagger}\right)\right]. 
\label{3DN4_S2}
\eea
Integrating out the auxiliary fields $\vec{H}(x)$ and combining the result with $\Delta S$ leads the term 
$\frac{1}{8g_0^2}\sum_x\tr \left[\vec{\Phi}'(x)+\Delta\vec{\Phi}'(x)\right]^2$. 
According to the result of appendix B, the minimum $\vec{\Phi}'(x)+\Delta\vec{\Phi}'(x)=0$ 
is given by the configurations satisfying  
\bea
 & & U_{23}(x)=U_{31}(x)=U_{12}(x)=1, \label{first} \\ 
 & & U_{4, -1}(x)=U_{4, -2}(x)=U_{4, -3}(x)=1. \label{second}
\eea
On the infinite lattice $x\in {\bf Z}^3$, eqs. (\ref{first}) are equivalent to  
$U_1(x)=U_2(x)=U_3(x)=1$ up to gauge transformations, and then eqs. (\ref{second}) 
mean $V(x)=V_0$ ($x$-independent constant). Furthermore, minimizing the term 
\beq
\tr \left(\phi(x)-V(x)\phi(x)V(x)^{\dagger}\right)
\left(\bar{\phi}(x)-V(x)\bar{\phi}(x)V(x)^{\dagger}\right) \nn 
\eeq
in the action 
for arbitrary $\phi(x)$, $\bar{\phi}(x)$ requires $V_0\in {\bf Z}_N$, i.e. a center element of SU($N$). 
Since all the ${\bf Z}_N$ elements for $V(x)=V_0$ give the identical value to the action, 
it is sufficient to consider the expansion of $V(x)$ around a representative element of ${\bf Z}_N$, 
say $1_N$. 
(If all the elements are taken into account, 
we end up merely with the overall factor $N$ multiplied to the 
partition function, which gives no influence on physics of the system.)   
Thus, we can justify the weak field expansion  
\beq
U_{\mu}(x) =e^{iaA_{\mu}(x)}=1+iaA_{\mu}(x)+\cdots, \qquad 
V(x)=e^{iav(x)}=1+iav(x)+\cdots. 
\label{expansion_3DN4}
\eeq

\paragraph{About Surplus Modes}
As argued in section 2.2, the fermion kinetic term in this three-dimensional model 
has no surplus mode, which obviously means fermion doublers not appearing. 
Then the exact supersymmetry $Q$ guarantees nonexistence of the surplus modes in the bosonic sector. 
Now we explicitly check it. 
After the weak field expansion (\ref{expansion_3DN4}), 
the Fourier transformation of $\vec{\Phi}'(x)$ becomes 
\bea
\tilde{\Phi}'_1(q) & = & 2a \left[(1-e^{-iaq_1})\tilde{v}(q) 
+(e^{iaq_2}-1)\tilde{A}_3(q) -(e^{iaq_3}-1)\tilde{A}_2(q)\right] 
+O(\tilde{v}^2, \tilde{A}^2), \nn \\
\tilde{\Phi}'_2(q) & = & 2a \left[(1-e^{-iaq_2})\tilde{v}(q) 
+(e^{iaq_3}-1)\tilde{A}_1(q) -(e^{iaq_1}-1)\tilde{A}_3(q)\right] 
+O(\tilde{v}^2, \tilde{A}^2), \nn \\
\tilde{\Phi}'_3(q) & = & 2a \left[(1-e^{-iaq_3})\tilde{v}(q) 
+(e^{iaq_1}-1)\tilde{A}_2(q) -(e^{iaq_2}-1)\tilde{A}_1(q)\right] 
+O(\tilde{v}^2, \tilde{A}^2). 
\eea
We do not have to take into account $\Delta\vec{\Phi}'(x)$,  
because its weak field expansion 
merely yields irrelevant interaction terms and does not contribute to the kinetic term. 

We can show that there is no nontrivial mode satisfying 
\beq
\tilde{\Phi}'_1(q)= \tilde{\Phi}'_2(q)= \tilde{\Phi}'_3(q)=0 + O(\tilde{v}^2, \tilde{A}^2)
\label{bosonic_surplus}
\eeq
except $q_1=q_2=q_3=0$. 
First, let us consider the cases that the momentum $(q_1, q_2, q_3)$ has only one nonzero component. 
Since eqs. (\ref{bosonic_surplus}) are invariant under the simultaneous permutations 
$(q_1, q_2, q_3) \limit (q_2, q_3, q_1)$ and 
$(\tilde{A}_1, \tilde{A}_2, \tilde{A}_3) \limit (\tilde{A}_2, \tilde{A}_3, \tilde{A}_1)$, 
it is sufficient to see the case $q_1=q_2=0, \, q_3\neq 0$, alone. 
Then, eqs. (\ref{bosonic_surplus}) lead $\tilde{v}=\tilde{A}_1=\tilde{A}_2=0$, 
which can be written as 
\beq
\tilde{v}^\alpha(q)=0, \quad 
\tilde{A}_{\mu}^\alpha(q)=(e^{iaq_{\mu}}-1)K^\alpha(q)
\label{gauge_dof}
\eeq
where $\alpha$ represents the index with respect to the SU($N$) gauge generators, 
and $K^\alpha(q)$ are complex-valued functions 
of $q=(0,0,q_3)$. 
It is not physically nontrivial, 
because it is a longitudinal mode and can be absorbed by the gauge degrees of freedom.  

Also, for the cases that the momentum has two nonzero components, 
the solution is described as the form (\ref{gauge_dof}) again
and it is not physically nontrivial. 
Finally, in the case of all the components of the momentum nonzero, 
under the gauge choice of $\tilde{A}_1(q)=0$ one can write $\vec{\tilde{\Phi}'}$ as 
\beq
\vec{\tilde{\Phi}'}(q)= M(q)\left(\begin{array}{c} 
\tilde{v}(q) \\ \tilde{A}_2(q) \\ \tilde{A}_3(q) \end{array} \right)
\eeq
with 
\beq
M(q)\equiv \left[\begin{array}{ccc} 
1-e^{-iaq_1} & 1-e^{iaq_3} & e^{iaq_2}-1 \\
1-e^{-iaq_2} & 0 & 1- e^{iaq_1} \\
1-e^{-iaq_3} & e^{iaq_1}-1 & 0 \end{array} \right]. 
\eeq
The determinant of the matrix $M(q)$ takes the form 
\beq
\det M(q) = -4(e^{iaq_1}-1)\left[
\sin^2 \left(\frac{aq_1}{2}\right) + \sin^2 \left(\frac{aq_2}{2}\right) 
+ \sin^2 \left(\frac{aq_3}{2}\right)\right]. 
\eeq
Since $\det M(q)\neq 0$ for $q_1, q_2, q_3\neq 0$, 
there is no nontrivial solution for (\ref{bosonic_surplus}). 

Thus, the boson kinetic term has no physical surplus modes, 
which precisely coincides with the situation 
for the fermionic sector as expected by the existence of the exact supersymmetry $Q$. 

\paragraph{Renormalization}
After the rescaling 
\bea
 & & \phi(x) \limit a\phi(x), \quad \bar{\phi}(x) \limit a \bar{\phi}(x), \quad 
\vec{H}(x) \limit a^2\vec{H}(x), \quad  
\psi_{\mu}(x) \limit a^{3/2}\psi_{\mu}(x), \nn \\
 & & \psi_4(x) \limit a^{3/2}\psi_4(x), \quad 
\vec{\chi}(x) \limit a^{3/2}\vec{\chi}(x), \quad \eta(x) \limit a^{3/2}\eta(x), 
\eea 
the lattice action (\ref{3DN4_S1}) reduces to the continuum action of ${\cal N}=4$ 
SYM in the continuum limit $a\limit 0$ with $g_3^{2}\equiv g_0^{2}/a$ fixed, 
and thus ${\cal N}=4$ supersymmetry and rotational symmetry 
in three dimensions are restored in the classical sense. 
The term $\Delta S'$ is of order $O(a^4)$ and becomes irrelevant. 
We will check whether the symmetry restoration persists against 
the quantum corrections, i.e. whether symmetries of the lattice action 
forbid relevant or marginal operators induced which possibly obstruct the symmetry 
restoration.  
In order to consider the quantum effects near the continuum limit, 
we assume the fixed point at $g_0=0$, which is suggested by the asymptotic freedom of the theory, 
and treat quantum fluctuations in the perturbative way around $g_0=0$.  

First let us consider the action $S^{\rm LAT}_{3D{\cal N}=4}$ without the term $\Delta S'$. 
It is useful for the renormalization argument 
to note symmetries of the lattice action (\ref{3DN4_S1}): 
\begin{itemize}
\item lattice translational symmetry 
\item SU($N$) gauge symmetry 
\item supersymmetry $Q$ 
\item global U$(1)_R$ internal symmetry\footnote{Because the theory is defined on odd-dimensional 
space-time, the U$(1)_R$ symmetry does not become anomalous.} 
of the same form as (\ref{U1R}) 
\item cyclic permutation 
$x\equiv (x_1, x_2, x_3) \limit \tilde{x}\equiv (x_3, x_1, x_2)$ as 
\bea
(U_1(x), U_2(x), U_3(x), V(x)) & \limit & 
(U_2(\tilde{x}), U_3(\tilde{x}), U_1(\tilde{x}), V(\tilde{x})) \nn \\
(\psi_1(x), \psi_2(x), \psi_3(x), \psi_4(x)) & \limit & 
(\psi_2(\tilde{x}), \psi_3(\tilde{x}), \psi_1(\tilde{x}), \psi_4(\tilde{x})) \nn \\
(H_1(x), H_2(x), H_3(x)) & \limit & (H_2(\tilde{x}), H_3(\tilde{x}), H_1(\tilde{x})) \nn \\
(\chi_1(x), \chi_2(x), \chi_3(x)) & \limit & 
(\chi_2(\tilde{x}), \chi_3(\tilde{x}), \chi_1(\tilde{x})) \nn \\
(\phi(x), \bar{\phi}(x)) & \limit &  (\phi(\tilde{x}), \bar{\phi}(\tilde{x})) \nn \\
\eta(x) & \limit & \eta(\tilde{x}). 
\label{perm_3DN4}
\eea
\end{itemize}
As for the cyclic permutation, $x\pm\hat{1}$, $x\pm\hat{2}$, $x\pm\hat{3}$ are transformed to 
$\tilde{x}\pm\hat{2}$, $\tilde{x}\pm\hat{3}$, $\tilde{x}\pm\hat{1}$, respectively. 
Then, one can see that 
\bea
\Phi_1'(x) +\Delta\Phi_1'(x) & \limit & \Phi_2'(\tilde{x}) +\Delta\Phi_2'(\tilde{x}),   \nn \\
\Phi_2'(x) +\Delta\Phi_2'(x) & \limit & \Phi_3'(\tilde{x}) +\Delta\Phi_3'(\tilde{x}),   \nn \\
\Phi_3'(x) +\Delta\Phi_3'(x) & \limit & \Phi_1'(\tilde{x}) +\Delta\Phi_1'(\tilde{x}), 
\eea
and that the action is invariant under (\ref{perm_3DN4}). 

The mass dimension of the coupling constant squared $g_3^2$ is one. 
For a generic boson field $\varphi$ (except the auxiliary fields $\vec{H}$) 
and a fermion field $\psi$, the dimensions 
are 1 and 3/2 respectively.   
Thus, operators of the type $\varphi^a \del^b\psi^{2c}$ 
have the dimension $p\equiv a+b+3c$, where 
`$\del$' means the derivative with respect to a coordinate. 
{}From the dimensional analysis, it turns out that the operators receive the following 
radiative corrections up to some powers of possible logarithmic factors:    
\beq
\left(\frac{a^{p-4}}{g_3^2} + c_1 a^{p-3} + c_2 a^{p-2}g_3^2 + 
c_3 a^{p-1}g_3^4 + c_4 a^pg_3^6 + \cdots\right)
\int \dd^3x\,  \varphi^a \del^b \psi^{2c},  
\label{loop_correction_3d}
\eeq 
where the first term in the parentheses represents the contribution from the tree level, 
and the term containing the coefficient $c_{\ell}$ comes from the $\ell$-loop contribution. 
It is clear from $g_3^2$ playing the same role as the Planck constant $\hbar$ 
in the action (\ref{3DN4_S1}). From the formula (\ref{loop_correction_3d}), 
one can see that operators with $p \leq 4-\ell$ are relevant or marginal 
if they are induced at the $\ell$-loop level. 

Because we know that the lattice action reduces to the desired continuum SYM action 
at the classical level, we need to check operators with $p=0, 1, 2, 3$ which are allowed 
to be induced by the above symmetries of the lattice action. 
Operators with $p\leq 4$ are listed 
in Table \ref{tab:operators}. The identity operator corresponding to $p=0$ 
merely shifts the origin of the energy, 
which is not interesting to us.   
%
\TABLE[t]{
\begin{tabular}{|c|ccccc|}
\hline \hline
$p=a+b+3c$ & 
\multicolumn{5}{|c|}{$\varphi^a\del^b\psi^{2c}$}  \\ 
\hline
0 &     &          &  1       &       &    \\
1 &     &          & $\varphi$  &      &     \\
2 &     &          & $\varphi^2$ &     &     \\
3 &     &$\varphi^3$, & $\psi\psi$, & $\varphi\del\varphi$  &     \\
4 & $\varphi^4$, & $\varphi^2\del\varphi$, & $(\del\varphi)^2$, & 
   $\psi\del\psi$, & $\varphi\psi\psi$ \\
\hline \hline
\end{tabular}
  \caption{List of operators with $p\leq 4$. 
}
\label{tab:operators}
}
%
For the cases $p=1, 2$, the SU($N$) gauge invariance and the U$(1)_R$ symmetry 
allow the scalar mass operators $\tr (\phi\bar{\phi})$, $\tr v^2$, while they are 
forbidden by the supersymmetry $Q$. 

In the $p=3$ case, let us first consider operators of the $Q$-exact form: 
$Q{\cal O}$ as $Q$-invariant operators. Then ${\cal O}$ has the mass dimension 
$5/2$ and the U$(1)_R$ charge $-1$. Candidates of such ${\cal O}$ are 
$\tr(\psi_{\mu}\bar{\phi})$, $\tr(\psi_4\bar{\phi})$, $\tr(\chi_{\bA}v)$. 
Here, we do not have to take into account $\tr(\eta v)$, 
because 
$Q\, \tr(\eta v)= -Q\, \tr(\psi_4\bar{\phi})$. 
Due to the requirement of the permutation symmetry (\ref{perm_3DN4}), 
the candidates reduce to the following 
three combinations:
\beq
\sum_{\mu =1}^3\tr(\psi_{\mu}\bar{\phi}), \qquad \tr(\psi_4\bar{\phi}), 
\qquad \sum_{\bA =1}^3\tr(\chi_{\bA}v). 
\eeq  
The corresponding $p=3$ operators are  
\bea
 & & Q\, \sum_{\mu =1}^3\tr(\psi_{\mu}\bar{\phi}) = 
\sum_{\mu =1}^3\tr(i\bar{\phi}D_{\mu}\phi -\psi_{\mu}\eta), \qquad 
Q\, \tr(\psi_4\bar{\phi}) = \tr (-[\phi, \bar{\phi}]v+\eta\psi_4), \nn \\
 & & Q \, \sum_{\bA =1}^3\tr(\chi_{\bA}v) = \sum_{\bA =1}^3 \tr(H_{\bA} v-\chi_{\bA}\psi_4). 
\label{3op_3DN4}
\eea
Note that other than the fermion bilinears appearing in the three operators of (\ref{3op_3DN4}) 
there are no possible $\psi\psi$ terms fulfilling the symmetry requirements. 
($\tr(\psi_{\mu}\chi_{\bA})$ comes from $Q\, \tr(A_{\mu}\chi_{\bA})$, 
while it is not gauge invariant.) 
Because any $p=3$ operators consistent to the symmetries must contain $\psi\psi$ terms, 
we can say that all the $p=3$ operators satisfying the symmetries are given by (\ref{3op_3DN4}) and 
nothing more appears. 
It means that there is no nontrivial $Q$-cohomology element 
satisfying the symmetries of the lattice action, and we do not have to consider operators 
belonging to the nontrivial $Q$-cohomology. 
Thus, we can conclude that the three of (\ref{3op_3DN4}) are the operators 
of $p=3$ possibly radiatively generated.    

When taking into account the effect of the term $\Delta S'$, it appears via the vertices of $\Delta S'$ 
in the loop expansion. Since $\Delta S'$ is of the order $O(a^4)$, it is easy to see that the effect 
to (\ref{loop_correction_3d}) 
becomes irrelevant in the continuum limit. Thus, the conclusion of the renormalization given above 
is valid without any change, even when the effect of $\Delta S'$ is included. 

Therefore, the loop corrections are allowed to generate the three operators (\ref{3op_3DN4}) alone 
except the identity.  
In order to reach the desired supersymmetric continuum theory, we have to add counter terms for the 
operators and cancel them with the radiative corrections by fine-tuning.

\subsection{2D ${\cal N}=4$ Case}
A lattice model for ${\cal N}=4$ SYM in two dimensions is obtained as a result of the further dimensional 
reduction of (\ref{3DN4_S1}) with respect to $x_3$. In addition to the variable $V(x)$ sitting on the 
two-dimensional lattice site $x=(x_1, x_2)$, there appear new site variables $W(x)\equiv U_3(x)$, 
$W(x)^{\dagger}=U_{-3}(x)$. 
The action $S^{{\rm LAT}}_{2D{\cal N}=4} + \Delta S'$ takes the form 
\bea
S^{{\rm LAT}}_{2D{\cal N}=4} & = & Q\frac{1}{2g_0^2}\sum_x \, \tr\left[ 
\frac14 \eta(x)\, [\phi(x), \,\bar{\phi}(x)] -i\vec{\chi}(x)\cdot(\vec{\Phi}'(x) 
+ \Delta\vec{\Phi}'(x))
+\vec{\chi}(x)\cdot\vec{H}(x)\right. \nn \\
 & & \hspace{-0.8cm}
+i\sum_{\mu=1}^2\psi_{\mu}(x)\left(\bar{\phi}(x) - 
U_{\mu}(x)\bar{\phi}(x+\hat{\mu})U_{\mu}(x)^{\dagger}\right)\nn \\
 & & \left.\hspace{-0.8cm}+i\psi_{3}(x)\left(\bar{\phi}(x) - 
W(x)\bar{\phi}(x)W(x)^{\dagger}\right)
+i\psi_{4}(x)\left(\bar{\phi}(x) - 
V(x)\bar{\phi}(x)V(x)^{\dagger}\right)\right],  
\label{2DN4_S1}
\eea
where $\vec{\Phi'}$, $\Delta\vec{\Phi'}$ are given by (\ref{Phip}) and 
\bea
 & & U_{-\mu, 4}(x)=U_{\mu}(x-\hat{\mu})^{\dagger}V(x-\hat{\mu})U_{\mu}(x-\hat{\mu})V(x)^{\dagger} 
= U_{4, -\mu}(x)^{\dagger}, 
\nn \\  
 & & U_{\mu 3}(x)= U_{\mu}(x)W(x+\hat{\mu})U_{-\mu}(x+\hat{\mu})W(x)^{\dagger} 
=U_{3 \mu}(x)^{\dagger}, 
\nn \\
 & & U_{-3,4}(x) = W(x)^{\dagger}V(x)W(x)V(x)^{\dagger}=U_{4, -3}(x)^{\dagger} \qquad (\mu=1, 2).  
\label{rule_dimred2}
\eea
$\Delta S'$ is given by the same expression as in (\ref{Delta_S'}). 

The explicit form of the action is  
\bea
S^{{\rm LAT}}_{2D{\cal N}=4} & = & \frac{1}{2g_0^2}\sum_x \, \tr\left[
\frac14 [\phi(x), \,\bar{\phi}(x)]^2 + \vec{H}(x)\cdot\vec{H}(x) 
-i\vec{H}(x)\cdot(\vec{\Phi}'(x)+\Delta\vec{\Phi}'(x)) \right. \nn \\
 & & \hspace{-1cm}
+\sum_{\mu=1}^2\left(\phi(x)-U_{\mu}(x)\phi(x+\hat{\mu})U_{\mu}(x)^{\dagger}
\right)\left(\bar{\phi}(x)-U_{\mu}(x)\bar{\phi}(x+\hat{\mu})
U_{\mu}(x)^{\dagger}\right) \nn \\
 & & \hspace{-1cm} + \left(\phi(x)-W(x)\phi(x)W(x)^{\dagger}\right)
\left(\bar{\phi}(x)-W(x)\bar{\phi}(x)W(x)^{\dagger}\right) \nn \\
 & & \hspace{-1cm} + \left(\phi(x)-V(x)\phi(x)V(x)^{\dagger}\right)
\left(\bar{\phi}(x)-V(x)\bar{\phi}(x)V(x)^{\dagger}\right) \nn \\
 & & \hspace{-1cm} -\frac14 \eta(x)[\phi(x), \,\eta(x)] 
- \vec{\chi}(x)\cdot [\phi(x), \,\vec{\chi}(x)] 
\nn \\
 & & \hspace{-1cm}
-\sum_{\mu=1}^2\psi_{\mu}(x)\psi_{\mu}(x)\left(\bar{\phi}(x)  + 
U_{\mu}(x)\bar{\phi}(x+\hat{\mu})U_{\mu}(x)^{\dagger}\right) \nn \\
 & & \hspace{-1cm}
-\psi_3(x)\psi_3(x)\left(\bar{\phi}(x) + W(x)\bar{\phi}(x)W(x)^{\dagger}\right)
-\psi_4(x)\psi_4(x)\left(\bar{\phi}(x) + V(x)\bar{\phi}(x)V(x)^{\dagger}\right) \nn \\
 & & \hspace{-1cm} + i\vec{\chi}(x)\cdot Q(\vec{\Phi}'(x)  +\Delta\vec{\Phi}'(x)) 
-i\sum_{\mu=1}^2\psi_{\mu}(x)\left(\eta(x)-
U_{\mu}(x)\eta(x+\hat{\mu})U_{\mu}(x)^{\dagger}\right) \nn \\ 
 & &\hspace{-1cm} \left. -i\psi_3(x)\left(\eta(x)-W(x)\eta(x)W(x)^{\dagger}\right)
-i\psi_4(x)\left(\eta(x)-V(x)\eta(x)V(x)^{\dagger}\right)\right]. 
\label{2DN4_S2}
\eea
{}From the argument parallel to the three-dimensional ${\cal N}=4$ case, the action is minimized by 
the configurations gauge-equivalent to $U_{\mu}(x)=1$, $V(x)=V_0\in{\bf Z}_N$, $W(x)=W_0\in{\bf Z}_N$. 
Since all the elements of ${\bf Z}_N$ for $V_0$ and $W_0$ give the same value to the action, 
it is justified to consider the weak field expansion around 1 as 
\bea
 & & U_{\mu}(x) =e^{iaA_{\mu}(x)}=1+iaA_{\mu}(x)+\cdots, \qquad 
V(x)=e^{iav(x)}=1+iav(x)+\cdots, \nn \\
 & & W(x)=e^{iaw(x)}=1+iaw(x)+\cdots. 
\label{expansion_2DN4}
\eea 
It reproduces the continuum action of the theory having the full ${\cal N}=4$ supersymmetry in the 
classical continuum limit: $a\limit 0$ with the coupling constant squared  $g_2^2= g_0^2/a^2$ fixed.   

Because the three-dimensional ${\cal N}=4$ SYM model in the previous subsection 
does not have any surplus modes, 
obviously no surplus mode appears in the two-dimensional ${\cal N}=4$ case here 
obtained by the further dimensional reduction therefrom\footnote{
In fact, the fermion kinetic part of (\ref{2DN4_S2}) takes the same form to 
the model based on the `BTFT formulation' preserving the two supercharges given 
in ref. \cite{sugino2}. The latter clearly has no surplus modes 
because of the property (\ref{no_surplus}). It also proves the nonexistence of the surplus modes 
for (\ref{2DN4_S2}).}. 

The argument for the renormalization goes along similarly to the three-dimensional ${\cal N}=4$ case. 
Now the coupling squared $g_2^2$ has the mass dimension two, and generic operators of the type 
$\varphi^a\del^b\psi^{2c}$ with the dimension $p\equiv a+b+3c$ receive the loop corrections as 
\beq
\left(\frac{a^{p-4}}{g_2^2} + c_1 a^{p-2} + c_2 a^pg_2^2 + \cdots\right)
\int \dd^2x \, \varphi^a \del^b \psi^{2c},  
\label{loop_correction_2d}
\eeq
where the notations are similar as in (\ref{loop_correction_3d}). 
$\Delta S'$ is of the order $O(a^4)$, and becomes irrelevant in the renormalization argument. 
To seek possible relevant or marginal operators radiatively generated, 
let us see operators of $p=1, 2$, which are of the types $\varphi$ and $\varphi^2$ according to 
Table \ref{tab:operators}. Among them, there does not exist any operator satisfying 
the SU($N$) gauge invariance, the U$(1)_R$ symmetry and the supersymmetry $Q$. 
Thus, the loop corrections are not allowed to 
generate any relevant or marginal operators 
except the identity, which means restoration of the full ${\cal N}=4$ supersymmetry and 
rotational invariance in the continuum limit without any fine-tuning. 

The lattice action (\ref{2DN4_S1}) preserving one supercharge $Q$ is different from the formulation 
based on the BTFT discussed in \cite{sugino2}, where two supercharges $Q_{\pm}$ are maintained. 
Thus, we have two different lattice formulations describing the same target theory of two-dimensional 
${\cal N}=4$ SYM with no tuning of parameters.

\setcounter{equation}{0}
\section{Naive 4D ${\cal N}= 4$ Lattice Model and Its Slight Modification} 
\label{sec:4DN4}
Here, we discuss on a naive lattice model for four-dimensional 
${\cal N}=4$ SYM as well as its slightly modified version. Notations are same as in ref.~\cite{sugino}. 

\subsection{Naive Lattice Model for 4D ${\cal N}=4$}
Let us start from the naive lattice action for four-dimensional ${\cal N}=4$ SYM theory with two 
supercharges $Q_{\pm}$ preserved: 
\bea
S^{{\rm LAT}}_{4D{\cal N}=4} &  = & 
Q_+Q_-\frac{1}{2g_0^2}\sum_x\, \tr \left[
-i\vec{B}(x)\cdot(\vec{\Phi}(x) + \Delta\vec{\Phi}(x)) \right. \nn \\
 & & \hspace{2cm} - \frac13\sum_{{\bA},{\bB},{\bC}=1}^3\varepsilon_{\bA\bB\bC}\,B_{\bA}(x)\,
[B_{\bB}(x), \,B_{\bC}(x)]  \nn \\ 
 & & \hspace{2cm}\left. 
- \sum_{\mu=1}^4\psi_{+\mu}(x)\psi_{-\mu}(x)-
\vec{\chi}_+(x)\cdot\vec{\chi}_-(x) 
-\frac14\eta_+(x)\eta_-(x)\right], 
\label{4DN4_S}
\eea
where quantities with the arrows are three-component vectors, 
and $\vec{\Phi}(x)$, $\Delta\vec{\Phi}(x)$ are given by (\ref{DPhin1}). 
In the ${\cal N}=4$ SYM, we have real scalar fields $\vec{B}$, $C$ (hermitian matrices) 
in addition to the complex scalars $\phi$, $\bar{\phi}$. 
 Also, $\tilde{H}_{\mu}$, $\vec{H}$ are bosonic auxiliary fields. 
Fields $\psi_{+ \mu}$, $\psi_{- \mu}$, $\vec{\chi}_+$, $\vec{\chi}_-$, $\eta_+$, 
$\eta_-$ represent fermions.  
The $Q_{\pm}$ supersymmetry transforms the lattice fields as 
\bea
 & & Q_+U_{\mu}(x) = i\psi_{+\mu}(x)U_{\mu}(x), \nn \\
 & & Q_-U_{\mu}(x) = i\psi_{-\mu}(x)U_{\mu}(x), \nn \\
 & & Q_+\psi_{+\mu}(x) = i\psi_{+\mu}\psi_{+\mu}(x) 
  -i\left(\phi(x)-U_{\mu}(x)\phi(x+\hat{\mu})U_{\mu}(x)^{\dagger}\right), 
\nn \\
 & & Q_-\psi_{-\mu}(x) = i\psi_{-\mu}\psi_{-\mu}(x) 
  +i\left(\bar{\phi}(x)-U_{\mu}(x)\bar{\phi}(x+\hat{\mu})U_{\mu}(x)^{\dagger}
\right), \nn \\
 & & Q_-\psi_{+\mu}(x) = \frac{i}{2}
\left\{\psi_{+\mu}(x), \,\psi_{-\mu}(x)\right\} -\frac{i}{2}
\left(C(x)-U_{\mu}(x)C(x+\hat{\mu})U_{\mu}(x)^{\dagger}\right) 
-\tilde{H}_{\mu}(x), \nn \\
 & & Q_+\psi_{-\mu}(x) = \frac{i}{2}
\left\{\psi_{+\mu}(x), \,\psi_{-\mu}(x)\right\} -\frac{i}{2}
\left(C(x)-U_{\mu}(x)C(x+\hat{\mu})U_{\mu}(x)^{\dagger}\right) 
+\tilde{H}_{\mu}(x), \nn \\
 & & Q_+\tilde{H}_{\mu}(x) = -\frac12
\left[\psi_{-\mu}(x), \,\phi(x)+U_{\mu}(x)\phi(x+\hat{\mu})U_{\mu}(x)^{\dagger}
\right] \nn \\
 & & \hspace{2cm} 
+\frac14\left[\psi_{+\mu}(x), \, C(x) +U_{\mu}(x)C(x+\hat{\mu})
U_{\mu}(x)^{\dagger}\right] \nn \\
 & & \hspace{2cm} +\frac{i}{2}\left(\eta_+(x) 
-U_{\mu}(x)\eta_+(x+\hat{\mu})U_{\mu}(x)^{\dagger}\right) \nn \\
 & & \hspace{2cm}  
+\frac{i}{2}\left[\psi_{+\mu}(x), \,\tilde{H}_{\mu}(x)\right] 
+\frac14\left[\psi_{+\mu}(x)\psi_{+\mu}(x), \,\psi_{-\mu}(x)\right], \nn \\
 & & Q_-\tilde{H}_{\mu}(x) = -\frac12
\left[\psi_{+\mu}(x), \,\bar{\phi}(x)+U_{\mu}(x)\bar{\phi}(x+\hat{\mu})
U_{\mu}(x)^{\dagger}\right] \nn \\
 & & \hspace{2cm}
-\frac14\left[\psi_{-\mu}(x), \, C(x) +U_{\mu}(x)C(x+\hat{\mu})
U_{\mu}(x)^{\dagger}\right] \nn \\
 & & \hspace{2cm} -\frac{i}{2}\left(\eta_-(x) 
-U_{\mu}(x)\eta_-(x+\hat{\mu})U_{\mu}(x)^{\dagger}\right) \nn \\
 & & \hspace{2cm}
+\frac{i}{2}\left[\psi_{-\mu}(x), \,\tilde{H}_{\mu}(x)\right] 
-\frac14\left[\psi_{-\mu}(x)\psi_{-\mu}(x), \,\psi_{+\mu}(x)\right],  
\label{group_A}
\eea
\bea
 & & Q_+\vec{B}(x) = \vec{\chi}_+(x), \quad Q_+\vec{\chi}_+(x) = [\phi(x), \,\vec{B}(x)], 
\nn \\  
 & & Q_-\vec{B}(x) = \vec{\chi}_-(x), \quad 
Q_-\vec{\chi}_-(x) = -[\bar{\phi}(x), \,\vec{B}(x)], \nn \\
 & & Q_-\chi_{+\bA}(x)= \frac12[C(x), \,B_{\bA}(x)] -\frac12\,\varepsilon_{\bA\bB\bC}\, 
[B_{\bB}(x), \,B_{\bC}(x)] - H_{\bA}(x), \nn \\
 & & Q_+\chi_{-\bA}(x)= \frac12[C(x), \,B_{\bA}(x)] +\frac12\,\varepsilon_{\bA\bB\bC}\, 
[B_{\bB}(x), \,B_{\bC}(x)] +H_{\bA}(x), \nn \\
 & & Q_+H_{\bA}(x) = [\phi(x), \,\chi_{-\bA}(x)] +\frac12[B_{\bA}(x), \,\eta_+(x)] 
-\frac12[C(x), \,\chi_{+\bA}(x)] \nn \\
 & & \hspace{2cm} -\varepsilon_{\bA\bB\bC}\, [B_{\bB}(x), \,\chi_{+\bC}(x)], 
\nn \\
 & &  Q_-H_{\bA}(x) = [\bar{\phi}(x), \,\chi_{+\bA}(x)] -\frac12[B_{\bA}(x), \,\eta_-(x)] 
+\frac12[C(x), \,\chi_{-\bA}(x)] \nn \\
 & & \hspace{2cm} -\varepsilon_{\bA\bB\bC}\, [B_{\bB}(x), \,\chi_{-\bC}(x)],  
\label{group_B}
\eea
\bea
 & & Q_+C(x) = \eta_+(x), \quad Q_+\eta_+(x) = [\phi(x), \,C(x)], \quad 
Q_-\eta_+(x) = -[\phi(x), \,\bar{\phi}(x)], \nn \\
 & & Q_-C(x) = \eta_-(x), \quad Q_-\eta_-(x) = -[\bar{\phi}(x), \,C(x)], \quad 
Q_+\eta_-(x) = [\phi(x), \,\bar{\phi}(x)], \nn \\
 & & Q_+\phi(x) = 0, \quad Q_-\phi(x)= -\eta_+(x), \quad 
Q_+\bar{\phi}(x) = \eta_-(x), \quad Q_-\bar{\phi}(x) = 0. 
\label{group_C}
\eea
The supercharges $Q_{\pm}$ are nilpotent up to gauge transformations in the sense 
\bea
Q_+^2 & = & 
(\mbox{infinitesimal gauge transformation with the parameter }\phi), \nn \\
Q_-^2 & = & 
(\mbox{infinitesimal gauge transformation with the parameter }-\bar{\phi}), 
\nn \\
\{Q_+, Q_-\} & = & 
 (\mbox{infinitesimal gauge transformation with the parameter }C). 
\label{nilpotent_N=4}
\eea
As discussed in ref. \cite{sugino}, 
the action has the ``double $Q$-exact form" ($Q_+Q_-(\mbox{something})$) 
based on the `BTFT formulation' \cite{btft} of ${\cal N}=4$ theories.   
The order of various quantities on the lattice are as follows: 
\bea
 & & \vec{B}(x), C(x), \phi(x), \bar{\phi}(x)=O(a), \quad 
\tilde{H}_{\mu}(x), \vec{H}(x) = O(a^2), \nn \\
 & & \psi_{\pm\mu}(x), \vec{\chi}_{\pm}(x), \eta_{\pm}(x) = O(a^{3/2}), \quad Q_{\pm} = O(a^{1/2}).
 \label{order_N4}
\eea
Note that the model has an SU$(2)_R$ symmetry, whose generators are defined as  
\bea
J_{++} & = & \sum_{x, \alpha} \left[
\sum_{\mu=1}^4 \psi_{+\mu}^\alpha(x)\frac{\del}{\del\psi_{-\mu}^\alpha(x)} + 
\sum_{\bA =1}^3 \chi_{+\bA}^\alpha(x)\frac{\del}{\del\chi_{-\bA}^\alpha(x)} - 
\eta_{+}^\alpha(x)\frac{\del}{\del\eta_{-}^\alpha(x)} \right. \nn \\
 & & \hspace{1cm} \left. + 2\phi^\alpha(x)\frac{\del}{\del C^\alpha(x)} 
- C^\alpha(x)\frac{\del}{\del \bar{\phi}^\alpha(x)}\right],  \nn \\
J_{--} & = & \sum_{x, \alpha} \left[
\sum_{\mu=1}^4 \psi_{-\mu}^\alpha(x)\frac{\del}{\del\psi_{+\mu}^\alpha(x)} + 
\sum_{\bA =1}^3 \chi_{-\bA}^\alpha(x)\frac{\del}{\del\chi_{+\bA}^\alpha(x)} - 
\eta_{-}^\alpha(x)\frac{\del}{\del\eta_{+}^\alpha(x)} \right. \nn \\
 & & \hspace{1cm} \left. - 2\bar{\phi}^\alpha(x)\frac{\del}{\del C^\alpha(x)} 
+ C^\alpha(x)\frac{\del}{\del \phi^\alpha(x)}\right],  \nn \\
J_0 & = & \sum_{x, \alpha} \left[
\sum_{\mu=1}^4 \psi_{+\mu}^\alpha(x)\frac{\del}{\del\psi_{+\mu}^\alpha(x)} 
-\sum_{\mu=1}^4 \psi_{-\mu}^\alpha(x)\frac{\del}{\del\psi_{-\mu}^\alpha(x)} 
+\sum_{\bA =1}^3 \chi_{+\bA}^\alpha(x)\frac{\del}{\del\chi_{+\bA}^\alpha(x)} \right. \nn \\
 & & \hspace{1cm}  -\sum_{\bA =1}^3 \chi_{-\bA}^\alpha(x)\frac{\del}{\del\chi_{-\bA}^\alpha(x)} 
+\eta_{+}^\alpha(x)\frac{\del}{\del\eta_{+}^\alpha(x)}
-\eta_{-}^\alpha(x)\frac{\del}{\del\eta_{-}^\alpha(x)}  \nn \\
 & & \hspace{1cm}\left.+2\phi^\alpha(x)\frac{\del}{\del\phi^\alpha(x)} 
-2\bar{\phi}^\alpha(x)\frac{\del}{\del\bar{\phi}^\alpha(x)}
\right]
\label{SU2R}
\eea
with $\alpha=1, \cdots, N^2-1$ the index of the SU($N$) gauge group generators, 
satisfying the algebra: 
\beq
[J_0, \, J_{++}] = 2J_{++}, \quad [J_0, \, J_{--}] = -2J_{--}, \quad 
[J_{++}, \, J_{--}] = J_0. 
\eeq
The SU$(2)_R$ symmetry is a subgroup of a full SU$(4)_R$ symmetry group 
in four-dimensional ${\cal N}=4$ SYM theory.  
$J_0$ is a generator of U$(1)_R$ rotation, 
which is contained in the SU$(2)_R$ as its Cartan subalgebra.  
$J_{++}$ ($J_{--}$) raises (lowers) the U$(1)_R$ charge by two-units. 
The subscript $\pm$ of the fermions indicates the U$(1)_R$ charge $\pm 1$, and 
$Q_{\pm}$ raises/lowers the charge by one. 
Under the SU$(2)_R$ rotation, each of $(\psi^\alpha_{+\mu}, \psi^\alpha_{-\mu})$, 
$(\chi^\alpha_+, \chi^\alpha_-)$, 
$(\eta^\alpha_+, -\eta^\alpha_-)$ and $(Q_+, Q_-)$ transforms as a doublet, 
and $(\phi^\alpha, C^\alpha, -\bar{\phi}^\alpha)$ as a triplet. 

Similarly to the case of the four-dimensional ${\cal N}=2$ model, 
we consider the action\footnote{We can consider to extend 
$\vec{B}(x)$, $\vec{\chi}_\pm(x)$, $\vec{H}(x)$ to hermitian matrices with nonvanishing trace parts. 
Then we have to soak up the would-be zero-modes of $\vec{B}^{(0)}(x)$, $\vec{\chi}^{(0)}_{\pm}(x)$. 
By a similar argument to the ${\cal N}=2$ case,  the effect is shown to be equivalent to adding the term 
$\Delta S$ (\ref{Delta_S}) to the action (\ref{4DN4_S}).} $S^{\rm LAT}_{4D{\cal N}=4} + \Delta S$.   
The minimum $\vec{\Phi}(x)+\Delta\vec{\Phi}(x)=0$ is given uniquely by $U_{\mu\nu}(x)=1$ 
for the parameter $r$ of (\ref{parameter_r}). 
Thus, the weak field expansion of $U_{\mu}(x)$ around 1 is justified again, 
but the same kind of surplus modes appear both in the boson and fermion 
kinetic terms. 
%
We will focus on constructing lower-dimensional lattice models for ${\cal N}=8$ SYM theories 
obtained by the dimensional reduction from a slightly modified version of the four-dimensional model. 

\subsection{Slightly Modified Model for 4D ${\cal N}=4$} 
Let us consider a slight modification of the four-dimensional model 
by replacing $\vec{\Phi}$, $\Delta\vec{\Phi}$ with 
$\vec{\Phi'}$, $\Delta\vec{\Phi'}$ defined in (\ref{Phip}). 
After rescaling the fermion fields by $a^{3/2}$, the fermion kinetic term becomes 
\bea
S^{(2)}_f & = & \frac{a^4}{2g_0^2}\sum_{x}\tr\left[\sum_{\mu =1}^4
\left\{-\frac12\Psi(x)^T\gamma_{\mu}(\Delta_{\mu}+\Delta^*_{\mu})\Psi(x) 
-a\frac12\Psi(x)^TP'_{\mu}\Delta_{\mu}\Delta^*_{\mu}\Psi(x) \right\}\right. \nn \\
  & & \hspace{2cm}
+ia\sum_{\bA =1}^3(\chi_{-\bA}(x)\Delta_4\Delta^*_{\bA}\psi_{+\bA}(x) 
-\psi_{+\bA}(x)\Delta_{\bA}\Delta^*_4\chi_{-\bA}(x) ) \nn \\
 & & \left. \frac{}{} \hspace{1.8cm}-ia\sum_{\bA =1}^3(\chi_{+\bA}(x)\Delta_4\Delta^*_{\bA}\psi_{-\bA}(x) 
-\psi_{-\bA}(x)\Delta_{\bA}\Delta^*_4\chi_{+\bA}(x) )
\right],   
\label{4DN4_Sf_modified}
\eea
with the explicit form of $P'_{\mu}$ in appendix C.3 obeying the algebra same as 
(\ref{gamma_P_algebra}).  
The fermion fields were combined as 
\beq
\Psi^T=\left(\psi_{+1}, \cdots, \psi_{+4},-\chi_{+1},\chi_{+2}, \chi_{+3}, \frac12\eta_+, 
\psi_{-1}, \cdots, \psi_{-4},-\chi_{-1},\chi_{-2}, \chi_{-3}, \frac12\eta_-\right). 
\eeq
The situation is parallel to the four-dimensional ${\cal N}=2$ case. 
Although the surplus modes still remain in the four-dimensional model even after the modification, 
they are removed by doing the dimensional reduction with respect to the fourth direction $x_4$. 
As a result, we obtain ${\cal N}=8$ SYM theory in three dimensions 
as well as that in two dimensions by reducing further.  
Thus, we can construct the ${\cal N}=8$ lattice models in two and three dimensions, 
which are {\it free from the surplus modes}.

\setcounter{equation}{0}
\section{Dimensionally Reduced ${\cal N}=8$ Models in Two and Three Dimensions}
\label{sec:23DN8}
\subsection{3D ${\cal N}=8$ Case}
After the dimensional reduction with respect to $x_4$, the slightly modified 
model in section 4.2 leads a lattice model for three-dimensional ${\cal N}=8$ SYM. 
The action $S^{{\rm LAT}}_{3D{\cal N}=8}+ \Delta S'$ is expressed as 
\bea
S^{{\rm LAT}}_{3D{\cal N}=8} &  = & 
Q_+Q_-\frac{1}{2g_0^2}\sum_x\, \tr \left[
-i\vec{B}(x)\cdot(\vec{\Phi'}(x) + \Delta\vec{\Phi'}(x)) \frac{}{} \right. \nn \\
 & & \hspace{2cm} - \frac13\sum_{{\bA},{\bB},{\bC}=1}^3\varepsilon_{\bA\bB\bC}\,B_{\bA}(x)\,
[B_{\bB}(x), \,B_{\bC}(x)]  \nn \\ 
 & & \hspace{2cm}\left. 
- \sum_{\mu=1}^4\psi_{+\mu}(x)\psi_{-\mu}(x)-
\vec{\chi}_+(x)\cdot\vec{\chi}_-(x) 
-\frac14\eta_+(x)\eta_-(x)\right], 
\label{3DN8_S}
\eea 
where $x$ represents a three-dimensional lattice site $(x_1, x_2, x_3)$, and 
$V(x)\equiv U_4(x)$, $V(x)^{\dagger}\equiv U_{-4}(x)$ become site variables. 
$\vec{\Phi'}(x)$, $\Delta\vec{\Phi'}(x)$ are given by 
the definition (\ref{Phip}) and (\ref{rule_dimred}). 
$\Delta S'$ is expressed as (\ref{Delta_S'}). 
The supersymmetry transformation (\ref{group_A}), (\ref{group_B}), (\ref{group_C}) 
is also dimensionally reduced with respect to $x_4$. 
The result changes the form of the $\mu=4$ part of (\ref{group_A}) alone. It becomes  
\bea
 & & Q_+V(x) = i\psi_{+4}(x)V(x), \nn \\
 & & Q_-V(x) = i\psi_{-4}(x)V(x), \nn \\
 & & Q_+\psi_{+4}(x) = i\psi_{+4}\psi_{+4}(x) 
  -i\left(\phi(x)-V(x)\phi(x)V(x)^{\dagger}\right), 
\nn \\
 & & Q_-\psi_{-4}(x) = i\psi_{-4}\psi_{-4}(x) 
  +i\left(\bar{\phi}(x)-V(x)\bar{\phi}(x)V(x)^{\dagger}
\right), \nn \\
 & & Q_-\psi_{+4}(x) = \frac{i}{2}
\left\{\psi_{+4}(x), \,\psi_{-4}(x)\right\} -\frac{i}{2}
\left(C(x)-V(x)C(x)V(x)^{\dagger}\right) 
-\tilde{H}_4(x), \nn \\
 & & Q_+\psi_{-4}(x) = \frac{i}{2}
\left\{\psi_{+4}(x), \,\psi_{-4}(x)\right\} -\frac{i}{2}
\left(C(x)-V(x)C(x)V(x)^{\dagger}\right) 
+\tilde{H}_4(x), \nn \\
 & & Q_+\tilde{H}_4(x) = -\frac12
\left[\psi_{-4}(x), \,\phi(x)+V(x)\phi(x)V(x)^{\dagger}
\right] \nn \\
 & & \hspace{2cm} 
+\frac14\left[\psi_{+4}(x), \, C(x) +V(x)C(x)V(x)^{\dagger}\right] \nn \\
 & & \hspace{2cm} +\frac{i}{2}\left(\eta_+(x) 
-V(x)\eta_+(x)V(x)^{\dagger}\right) \nn \\
 & & \hspace{2cm}  
+\frac{i}{2}\left[\psi_{+4}(x), \,\tilde{H}_4(x)\right] 
+\frac14\left[\psi_{+4}(x)\psi_{+4}(x), \,\psi_{-4}(x)\right], \nn \\
 & & Q_-\tilde{H}_4(x) = -\frac12
\left[\psi_{+4}(x), \,\bar{\phi}(x)+V(x)\bar{\phi}(x)V(x)^{\dagger}\right] \nn \\
 & & \hspace{2cm}
-\frac14\left[\psi_{-4}(x), \, C(x) +V(x)C(x)V(x)^{\dagger}\right] \nn \\
 & & \hspace{2cm} -\frac{i}{2}\left(\eta_-(x) 
-V(x)\eta_-(x)V(x)^{\dagger}\right) \nn \\
 & & \hspace{2cm}
+\frac{i}{2}\left[\psi_{-4}(x), \,\tilde{H}_4(x)\right] 
-\frac14\left[\psi_{-4}(x)\psi_{-4}(x), \,\psi_{+4}(x)\right].  
\label{3DN4_group_A}
\eea

Repeating the argument similar to the three-dimensional ${\cal N}=4$ case, we can justify the weak 
field expansion (\ref{expansion_3DN4}) 
and show nonexistence of the surplus modes.  

\paragraph{Renormalization}
After the rescaling 
\bea
 & & \phi(x) \limit a\phi(x), \quad \bar{\phi}(x) \limit a\bar{\phi}(x), 
\quad \vec{B}(x)\limit a\vec{B}(x), \quad C(x)\limit a C(x), \nn \\
 & & \tilde{H}_{\mu}(x) \limit a^2\tilde{H}_{\mu}(x), \quad \tilde{H}_4(x)\limit a^2\tilde{H}_4(x), 
\quad \vec{H}(x)\limit a^2\vec{H}(x), \nn \\
 & & \psi_{\pm\mu}(x)\limit a^{3/2}\psi_{\pm\mu}(x), 
\quad \psi_{\pm 4}(x)\limit a^{3/2}\psi_{\pm 4}(x), 
\quad \vec{\chi}_{\pm}(x)\limit a^{3/2}\vec{\chi}_{\pm}(x), \nn \\
 & & \eta_{\pm}(x) \limit a^{3/2}\eta_{\pm}(x), 
\eea  
the classical action (\ref{3DN8_S}) reduces to the desired continuum action for 
${\cal N}=8$ SYM in three dimensions 
in the continuum limit $a\limit 0$ with $g_3^2\equiv g_0^2/a$ fixed. 
Then, ${\cal N}=8$ supersymmetry and rotational symmetry are restored. 
Now, 
we argue about the renormalization with noting symmetries of the lattice 
action (\ref{3DN8_S}): 
\begin{itemize}
\item lattice translational symmetry 
\item SU($N$) gauge symmetry 
\item supersymmetry $Q_{\pm}$
\item SU$(2)_R$ symmetry of the same form as (\ref{SU2R})
\item interchanging symmetry of $Q_+\leftrightarrow Q_-$ with 
\bea
 & & \phi(x)\limit -\bar{\phi}(x), 
\quad \bar{\phi}(x)\limit -\bar{\phi}(x), \quad \eta_{\pm}(x)\limit\eta_{\mp}(x), 
\nn \\
 & & 
\tilde{H}_{\mu}(x)\limit -\tilde{H}_{\mu}(x), \quad \tilde{H}_4(x) \limit -\tilde{H}_4(x), \nn \\
 & & \psi_{\pm\mu}(x) \limit \psi_{\mp\mu}(x), \quad \psi_{\pm 4}(x) \limit \psi_{\mp 4}(x), 
\quad \vec{B}(x)\limit -\vec{B}(x), \quad \vec{\chi}_{\pm}(x) \limit -\vec{\chi}_{\mp}(x), \nn \\
 & & \mbox{others ($U_{\mu}(x)$, $V(x)$, $\vec{H}(x)$, $C(x)$) not changed.}
\label{Qpm_symmetry} 
\eea
\item cyclic permutation $x\equiv (x_1, x_2, x_3)\limit \tilde{x}\equiv (x_3, x_1, x_2)$ as 
\bea
(U_1(x), U_2(x), U_3(x), V(x)) & \limit & (U_2(\tilde{x}), U_3(\tilde{x}), U_1(\tilde{x}), V(\tilde{x})) 
\nn \\
(\psi_{\pm 1}(x), \psi_{\pm 2}(x), \psi_{\pm 3}(x), \psi_{\pm 4}(x)) & \limit & 
(\psi_{\pm 2}(\tilde{x}), \psi_{\pm 3}(\tilde{x}), \psi_{\pm 1}(\tilde{x}), \psi_{\pm 4}(\tilde{x}))
\nn \\
(\tilde{H}_1(x), \tilde{H}_2(x), \tilde{H}_3(x), \tilde{H}_4(x)) & \limit & 
(\tilde{H}_2(\tilde{x}), \tilde{H}_3(\tilde{x}), \tilde{H}_1(\tilde{x}), \tilde{H}_4(\tilde{x}))
\nn \\
(B_1(x), B_2(x), B_3(x)) & \limit & (B_2(\tilde{x}), B_3(\tilde{x}), B_1(\tilde{x})) \nn \\
(\chi_{\pm 1}(x), \chi_{\pm 2}(x), \chi_{\pm 3}(x)) & \limit & 
(\chi_{\pm 2}(\tilde{x}), \chi_{\pm 3}(\tilde{x}), \chi_{\pm 1}(\tilde{x})) \nn	 \\
(H_1(x), H_2(x), H_3(x)) & \limit & (H_2(\tilde{x}), H_3(\tilde{x}), H_1(\tilde{x})) \nn \\
(\phi(x), \bar{\phi}(x), C(x)) & \limit & (\phi(\tilde{x}), \bar{\phi}(\tilde{x}), C(\tilde{x})) \nn \\
\eta_{\pm}(x) & \limit & \eta_{\pm}(\tilde{x}). 
\label{3DN8_perm}
\eea
\end{itemize}
Parallel to the argument given in the three-dimensional ${\cal N}=4$ case,  
operators we have to check are those with the dimension $p=1, 2, 3$ of the types appearing 
in Table \ref{tab:operators}. 
The effect of $\Delta S'$ is irrelevant in the argument of the renormalization. 
For $p=1, 2$, the four operators $\tr(4\phi\bar{\phi}+C^2)$, $\tr v^2$, $\tr(B_{\bA}v)$, 
$\tr(B_{\bA}B_{\bB})$ satisfy the requirement from 
the SU($N$) gauge invariance and the SU$(2)_R$ symmetry, 
but all of them conflict with the supersymmetry $Q_{\pm}$. 

Next, to search the $p=3$ operators, it is convenient to utilize the result of $Q_{\pm}$-cohomology. 
Under the assignment of the degrees $(q_+, q_-)$ to each variable: 
\beq
\begin{array}{cc}
U_{\mu}: & (0,0) \\
\psi_{+\mu}: & (1,0) \\
\psi_{-\mu}: & (0, 1) \\
\tilde{H}_{\mu}: & (1,1) 
\end{array}
\qquad 
\begin{array}{cc}
\vec{B}: & (1,1) \\
\vec{\chi}_{+}: & (2,1) \\
\vec{\chi}_{-}: & (1, 2) \\
\vec{H}: & (2,2) 
\end{array}
\qquad 
\begin{array}{cc}
\phi: & (2,0) \\
\bar{\phi}: & (0,2) \\
C: & (1, 1) \\
\eta_+: & (2,1) \\
\eta_-: & (1,2) 
\end{array}
\eeq
(here, $\mu=1, \cdots, 4$), the result is summarized as follows:
\\
\noindent 
{\it Any gauge invariant and SU$(2)_R$ invariant operator $\alpha$ annihilated 
by $Q_{\pm}$ can be expressed as   
\bea
\alpha & = &  \alpha_0 + Q_+\beta_- + Q_+Q_-\gamma \nn \\
       & = &  \alpha_0 + Q_-\beta_+ + Q_+Q_-\gamma. 
\label{result_cohomology}
\eea
$\alpha_0$ is proportional to the identity, and $\beta_{\pm}$, $\gamma$ are 
gauge invariant. 
$\beta_{+}$ ($\beta_-$) is a $Q_{+}$- ($Q_-$-) closed operator of
the degrees $(1,0)$ ($(0,1)$), and has nontrivial cohomology with 
respect to $Q_{+}$ ($Q_-$).} (For a proof of this statement, see appendix C in ref. \cite{sugino}.) 
Since the mass dimension of $\beta_{\pm}$ is $5/2$, gauge invariant candidates for $\beta_-$ are 
$\tr(v\psi_{-\mu})$ alone. However, they are not $Q_-$ invariant. 
Hence, there is no operator of the type $Q_+\beta_-$. 
Similar holds for $Q_-\beta_+$ from the $Q_+\leftrightarrow Q_-$ 
symmetry. $\gamma$ has the dimension two with the U$(1)_R$ neutral, 
and it must be odd under $Q_+\leftrightarrow Q_-$ from the symmetry requirement (\ref{Qpm_symmetry}). 
It leads the six candidates $\tr(B_{\bA}v)$, $\tr(B_{\bA}C)$ 
($\bA =1, 2, 3$) for $\gamma$. Among them, the two combinations 
$\tr[(B_1+B_2+B_3)v]$, $\tr[(B_1+B_2+B_3)C]$ survive by imposing the requirement (\ref{3DN8_perm}). 
Furthermore, the SU$(2)_R$ invariance of $Q_+Q_-\gamma$ singles out the former. 
Thus, we can conclude that the $p=3$ operator 
\bea
Q_+Q_-\tr[(B_1+B_2+B_3)v] & = & \sum_{\bA =1}^3\tr\left[\left(\frac12[C, B_{\bA}]+H_{\bA}
+\sum_{\bB, \bC =1}^3\frac12\varepsilon_{\bA\bB\bC}[B_{\bB}, B_{\bC}] 
\right)v\right. \nn \\
 & & \left. +B_{\bA}\left(-\frac12[v, C]+\tilde{H}_4\right)
  +\chi_{+\bA}\psi_{- 4}-\chi_{-\bA}\psi_{+4}\right],
\eea
which is the only one consistent to the symmetries possessed by the lattice action, 
is possibly generated as the radiative corrections. 
In order to reach the desired continuum theory with the full ${\cal N}=8$ supersymmetry restored, 
it is necessary to add a counter term corresponding 
the above operator to cancel it with the radiative corrections by fine-tuning.

\subsection{2D ${\cal N}=8$ Case}
A lattice theory for ${\cal N}=8$ SYM in two dimensions is obtained by 
the further dimensional reduction from the three-dimensional ${\cal N}=8$ model (\ref{3DN8_S}) 
with respect to $x_3$. 
The action takes the same form as (\ref{3DN8_S})$+\Delta S'$, but now $x$ stands for a two-dimensional lattice site 
$(x_1, x_2)$. Also, the variables $W(x)\equiv U_3(x)$, $W(x)^{\dagger}=U_{-3}(x)$ become sitting at sites. 
$\vec{\Phi'}$, $\Delta\vec{\Phi'}$ are defined by (\ref{Phip}) and (\ref{rule_dimred2}). 

When considering fluctuations around the minimum of the action, the weak field expansion 
of the form (\ref{expansion_2DN4}) 
is justified, and no surplus mode appears. 
The naive continuum limit $a\limit 0$ with $g_2^2\equiv g_0^2/a^2$ fixed 
leads the desired continuum action after a suitable rescaling of lattice fields corresponding to 
the order (\ref{order_N4}). Because any operators of the types $\varphi$, $\varphi\varphi$ with $p=1, 2$ 
in Table \ref{tab:operators} 
do not satisfy the requirement from the $Q_{\pm}$ supersymmetry, 
the loop corrections are not allowed to generate 
any relevant or marginal operators except the identity. 
It means restoration of the full supersymmetry and 
the rotational invariance in the continuum limit at the quantum level without fine-tuning.

\setcounter{equation}{0}
\section{Summary and Discussions}
\label{sec:summary}
In this paper, first we have discussed on naive lattice models for ${\cal N}=2, \, 4$ SYM theories 
in four-dimensions constructed based on the `TFT, BTFT formulations' along the same line 
to the papers~\cite{sugino,sugino2}. 
A problem of the degenerate vacua encountered in the paper \cite{sugino} 
is resolved by extending some fields and soaking up would-be zero-modes in the continuum limit. 
We see that 
there appear surplus zero-modes carrying nonzero momenta in the kinetic terms for both of bosons and 
fermions 
in addition to the ordinary zero-momentum mode. 
The appearence of the surplus modes seems to be related to the fact that 
our construction of a lattice counter part 
of the quantity $\int F\wedge F$ is not exactly topological on the lattice. 

By dimensional reduction of a slightly modified version of the four-dimensional models, 
we have constructed two- and three-dimensional models with ${\cal N}=4, \, 8$ supersymmetry, 
{\it which are completely free from the surplus modes}. These models flow into the desired continuum 
theories with three and one parameters fine-tuned for three-dimensional ${\cal N}=4, \, 8$ 
cases, while no fine-tuning is required in the two-dimensional cases. 
This is a main result of this paper. 

After building the models, it is natural to consider actual numerical computation by utilizing 
the lattice actions presented here. A first step toward the end 
will be to examine the positivity of the fermion 
determinant similarly as in refs. \cite{giedt} 
for the case of lattice models based on the idea of the deconstruction \cite{kaplan2}. 
Also, for the numerical study, 
it would be attractive to apply the method of density matrix renormalization group \cite{sugihara}
to our lattice models via the Hamiltonian formulation. 


As for four-dimensional ${\cal N}=4$ theory, 
it is believed that it has no anomaly due to the ultra-violet finiteness \cite{mandelstam}  
differently from the ${\cal N}=2$ case. 
Suppose that we may not consider about the problem of the surplus modes. 
Then, since the supersymmetry $Q_{\pm}$ of 
the ${\cal N}=4$ lattice model 
enhances to the full ${\cal N}=4$ supersymmetry in the naive continuum limit, 
only the lattice artifact of $O(a)$ breaks the full 16 supercharges down to the two $Q_{\pm}$ 
and violates the ultra-violet finiteness. It induces loop divergence 
as $a\limit 0$. Thus, the radiative corrections seem to have the structure something like 
\beq
(\mbox{lattice artifact of $O(a)$})\times  (\mbox{loop divergence as $a\limit 0$}),  \nn
\eeq
which are somewhat subtle to say whether they survive in the continuum limit or not.    
It would be interesting to do an explicit perturbative analysis for the 
four-dimensional ${\cal N}=4$ lattice model, 
and to see whether the radiative corrections induce relevant or marginal 
operators to the action. 
 
${\cal N}=8$ SYM theories in two and three dimensions 
have been proposed as matrix string models which serve 
nonperturbative definitions of superstring theory \cite{DVVBS}. 
It would be meaningful to do nonperturbative investigation of the corresponding lattice models 
presented here from the viewpoint of string theory as well as the field-theoretical interest. 
  
%

\acknowledgments
The author would like to thank S.~Catterall, K.~Fujikawa, J.~Giedt, 
M.~Kato, I.~Kanamori, N.~Kawamoto, 
Y.~Kikukawa, K.~Nagata, M.~Sakamoto, H.~So, T. Sugihara and T.~Takimi   
for variable discussions and useful conversations.


\appendix
\section{On Solutions of Eq. (\ref{vacuaSU(N)})}
\label{sec:vacSU(N)}
\setcounter{equation}{0}
\renewcommand{\theequation}{A.\arabic{equation}}
In this appendix, we show that solutions of the equation for $U$, $V\in {\rm SU}(N)$ 
\beq
e^{-i\varphi}(U+V) + e^{i\varphi}(U^\dagger +V^\dagger) = 
 \left(\frac1N \,\tr\left[e^{-i\varphi}(U+V) + e^{i\varphi}(U^\dagger +V^\dagger)\right]\right){\bf 1}_N
\label{eqUV_SU}
\eeq
are not unique and distribute continuously from $U=V=1$. 
It leads that the solutions of (\ref{vacuaSU(N)}) also distribute continuously 
from $U_{\mu\nu}(x) =1$. 

Diagonalizing $U, V$ as 
\beq
U=\Omega_U\left( \begin{array}{ccc} e^{iu_1} & & \\  & \ddots &  \\  & & e^{iu_N}\end{array}
\right) \Omega_U^\dagger, \qquad 
V=\Omega_V\left( \begin{array}{ccc} e^{iv_1} & & \\  & \ddots &  \\  & & e^{iv_N}\end{array}
\right) \Omega_V^\dagger 
\eeq
with $\Omega_U, \Omega_V\in \mbox{SU}(N)$, 
%
eq. (\ref{eqUV_SU}) can be expressed in the form 
\beq
\left( \begin{array}{ccc} \cos(u_1-\varphi) & & \\  & \ddots &  \\  & & \cos(u_N-\varphi)
\end{array} \right) + 
\Omega \left( \begin{array}{ccc} \cos(v_1-\varphi) & & \\  & \ddots &  \\  & & \cos(v_N-\varphi)
\end{array} \right) \Omega^\dagger = c{\bf1}_N 
\label{eqUV2_SU}
\eeq
with $c$ some real constant. 
Since $\Omega\equiv \Omega_U^\dagger\Omega_V$ is adjoint action transforming from a 
diagonal matrix to a diagonal matrix, it must belong to the Weyl group of SU$(N)$, i.e. 
$\Omega$ leads even permutations of diagonal entries. 
Thus eq. (\ref{eqUV2_SU}) is equivalent to the following set of equations: 
\beq
\left\{ \begin{array}{ccc}
\cos(u_1-\varphi) + \cos(v_{i(1)}-\varphi) & = & c,  \\
\vdots \hspace{1cm} & &  \vdots \\
\cos(u_N-\varphi) + \cos(v_{i(N)}-\varphi) & = & c, 
\end{array}\right. 
\label{eqUV3_SU}
\eeq
with an even permutation: $(1, \cdots, N)\rightarrow (i(1), \cdots, i(N))$.  

Let us take a parametrization 
\beq
u_1 + \cdots + u_N=0, \quad v_1+ \cdots + v_N=0. 
\label{uN_vN}
\eeq
Because the permutation is just renaming the variables, we may consider 
the case $i(1)=1, \cdots, i(N)=N$ without loss of generality. 
Then the equation can be written as  
\bea
\cos(u_1-\varphi) + \cos(v_1-\varphi) & = & \cos(u_2-\varphi) + \cos(v_2-\varphi), \nn \\
\cos(u_1-\varphi) + \cos(v_1-\varphi) & = & \cos(u_3-\varphi) + \cos(v_3-\varphi), \nn \\ 
  \vdots \hspace{2cm}   &   & \hspace{2cm}    \vdots  \nn \\
\cos(u_1-\varphi) + \cos(v_1-\varphi) & = & \cos(u_N-\varphi) + \cos(v_N-\varphi).
\label{eqUV4_SU}
\eea
Let us consider the case $u_1=\cdots = u_{N-1}\equiv u$, $v_1=\cdots = v_{N-1} \equiv v$. 
{}From (\ref{uN_vN}), $u_N=-(N-1)u$, $v_N = -(N-1)v$. 
Among (\ref{eqUV4_SU}), the equation 
\beq
f(u) = -f(v) 
\label{eqUV5_SU}
\eeq
with 
\beq
f(x) \equiv \cos(x-\varphi)- \cos((N-1)x+\varphi) 
\eeq
only remains nontrivial. 
When $|x|$ is small, $f(x)$ is monotonically increasing ($f'(x)>0$) for 
\beq
(N-1)|x|<\varphi <\pi-(N-1)|x|.
\label{varphi_range}
\eeq
Thus, there exist solutions of (\ref{eqUV5_SU}) continuously connecting from the trivial one $u=v=0$ 
for $\varphi$ satisfying (\ref{varphi_range}) with $|x|=\max (|u|, |v|)$. 

Suppose that we consider the case $U=U_{14}$, $V=U_{23}$ and $U_1=U_1(x_1, x_4)$ ($U_1$ depends only on $x_1$ and $x_4$), 
$U_2=U_2(x_2, x_3)$ ($U_2$ depends only on $x_2$ and $x_3$), 
$U_3=U_4=1$, for example. Then, $U_{24}(x)=U_{34}(x)=U_{12}(x)=U_{31}(x)=1$, and the equations 
\beq
\Phi_2(x)+\Delta\Phi_2(x)\propto {\bf 1}_N, \qquad 
\Phi_3(x)+\Delta\Phi_3(x)\propto {\bf 1}_N
\eeq
are trivially satisfied. The remaining $\Phi_1(x)+\Delta\Phi_1(x)\propto {\bf 1}_N$ has solutions 
continuously connecting from $U_{14}(x)=U_{23}(x)=1$. 

\section{Resolution of Vacuum Degeneracy}
\label{sec:vac}
\setcounter{equation}{0}
\renewcommand{\theequation}{B.\arabic{equation}}
Here we show that the minimum 
\beq
\vec{\Phi}(x)+\Delta\vec{\Phi}(x)=0 
\label{minimum}
\eeq 
of the 
gauge field action is uniquely given by the configuration $U_{\mu\nu}(x)=1$ 
when the parameter $r=\cot \varphi$ ranging as 
\beq
0<\varphi\leq \frac{\pi}{2N}.  
\label{varphi}
\eeq
To do that, it is sufficient to see that the equation for $U, V\in$ SU$(N)$: 
\beq
e^{-i\varphi}(-2+U+V) + e^{i\varphi}(-2+U^\dagger +V^\dagger)=0 
\label{eqUV}
\eeq
has the unique solution $U=V=1$ when $\varphi$ satisfies (\ref{varphi}). 

Diagonalizing $U, V$ as 
\beq
U=\Omega_U\left( \begin{array}{ccc} e^{iu_1} & & \\  & \ddots &  \\  & & e^{iu_N}\end{array}
\right) \Omega_U^\dagger, \qquad 
V=\Omega_V\left( \begin{array}{ccc} e^{iv_1} & & \\  & \ddots &  \\  & & e^{iv_N}\end{array}
\right) \Omega_V^\dagger 
\eeq
with $\Omega_U, \Omega_V\in \mbox{SU}(N)$, 
%
eq. (\ref{eqUV}) can be expressed in the form 
\beq
\Omega \left( \begin{array}{ccc} \cos(v_1-\varphi) & & \\  & \ddots &  \\  & & \cos(v_N-\varphi)
\end{array} \right) \Omega^\dagger = 
2\cos\varphi - 
\left( \begin{array}{ccc} \cos(u_1-\varphi) & & \\  & \ddots &  \\  & & \cos(u_N-\varphi)
\end{array} \right). 
\label{eqUV2}
\eeq
Since $\Omega\equiv \Omega_U^\dagger\Omega_V$ is adjoint action transforming from a 
diagonal matrix to a diagonal matrix, it must belong to the Weyl group of SU$(N)$, i.e. 
$\Omega$ leads even permutations of diagonal entries. 
Thus eq. (\ref{eqUV2}) is equivalent to the following set of equations: 
\beq
\left\{ \begin{array}{ccc}
\cos(v_{i(1)}-\varphi) & = & 2\cos\varphi - \cos(u_1-\varphi),  \\
\vdots \hspace{1cm} & &  \hspace{1cm}\vdots \\
\cos(v_{i(N)}-\varphi) & = & 2\cos\varphi - \cos(u_N-\varphi), 
\end{array}\right. 
\label{eqUV3}
\eeq
with an even permutation: $(1, \cdots, N)\rightarrow (i(1), \cdots, i(N))$. 

Let us consider the equations for an arbitrarily fixed permutation $i$. 
We may take the following parametrizations: 
\bea
& & -\pi < u_1, \cdots, u_{N-1} \leq \pi, \qquad 
u_N = -u_1-\cdots -u_{N-1}, \nn \\
& & -\pi < v_{i(1)}, \cdots, v_{i(N-1)} \leq \pi, \qquad 
v_{i(N)} = -v_{i(1)}-\cdots -v_{i(N-1)}. 
\label{parametrization_uv}
\eea 
Eq. (\ref{eqUV3}) can be written as  
\beq
\cos\left(\frac{u_k+v_{i(k)}}{2}-\varphi\right)\cos\left(\frac{u_k-v_{i(k)}}{2}\right)
=\cos\varphi \qquad (k=1, \cdots, N). 
\label{eqUV4}
\eeq
In eqs. (\ref{eqUV4}), 
$\left|\cos\left(\frac{u_k-v_{i(k)}}{2}\right)\right|\leq 1$ requires 
\beq
\left|\cos\left(\frac{u_k+v_{i(k)}}{2}-\varphi\right)\right|\geq\cos\varphi, 
\eeq
which means 
\beq
\pi n\leq \frac{u_k+v_{i(k)}}{2}\leq2\varphi+\pi n \qquad (n=0, \pm1, \cdots). 
\label{eqUV5}
\eeq
\paragraph{The case $k=1, \cdots, N-1$ in eq. (\ref{eqUV5})} 
First consider the case $k=1, \cdots, N-1$.  
{}From the parametrization (\ref{parametrization_uv}), the relation (\ref{eqUV5}) is realized 
by the following three cases: 
\bea
& & 0\leq \frac{u_k+v_{i(k)}}{2}\leq 2\varphi, \label{yoi} \\
& & \frac{u_k+v_{i(k)}}{2} = \pi, \label{dame1} \\
& & -\pi < \frac{u_k+v_{i(k)}}{2} \leq 2\varphi -\pi. \label{dame2}
\eea
The second possibility (\ref{dame1}) means $u_k=v_{i(k)}=\pi$, 
but it is not consistent to (\ref{eqUV4}). 
Also, the third (\ref{dame2}) leads $\cos\left(\frac{u_k+v_{i(k)}}{2}\right)<0$, which 
requires 
\beq
\cos\left(\frac{u_k-v_{i(k)}}{2}\right)<0 
\label{cos_u-v}
\eeq
from the consistency with eq. (\ref{eqUV4}). Then, we obtain 
\beq
\frac{\pi}{2}< \frac{u_k-v_{i(k)}}{2}<\pi
\qquad \mbox{or} \qquad 
-\pi < \frac{u_k-v_{i(k)}}{2}<-\frac{\pi}{2}
\label{u-v}
\eeq
from (\ref{cos_u-v}). However, it is easy to see that (\ref{dame2}) and (\ref{u-v}) are not 
compatible with each other. Thus, only the possibility (\ref{yoi}) remains. 

\paragraph{The case $k=N$ in eq. (\ref{eqUV5})} 
Next, we see the case $k=N$, which reads 
\beq
-2\varphi +\pi n \leq \frac12 (u_1+\cdots +u_{N-1} +v_{i(1)} +\cdots +v_{i(N-1)}) \leq \pi n
\label{u+v_1}
\eeq
with $n=0, \pm1, \cdots$. 
On the other hand, summing (\ref{yoi}) over $k=1, \cdots, N-1$ gives 
\beq
0\leq \frac12 (u_1+\cdots + u_{N-1}+v_{i(1)}+\cdots +v_{i(N-1)})\leq 2(N-1)\varphi. 
\label{u+v_2}
\eeq
{}From these two equations, we obtain 
\beq
\frac12(u_1+\cdots +u_{N-1} +v_{i(1)} +\cdots +v_{i(N-1)})=\left\{\begin{array}{ll}
0, \pi\frac{N-1}{N} & \quad \mbox{for }\varphi = \frac{\pi}{2N}. \\
0 & \quad \mbox{for } 0< \varphi < \frac{\pi}{2N}. \end{array}\right. 
\eeq  

In the case $\frac12(u_1+\cdots +u_{N-1} +v_{i(1)} +\cdots +v_{i(N-1)})=
\pi\frac{N-1}{N}$ for $\varphi=\frac{\pi}{2N}$, owing to (\ref{yoi}) we have 
\beq
u_1+v_{i(1)}= \cdots = u_{N-1}+v_{i(N-1)} = \frac{2\pi}{N}.   
\eeq
(\ref{eqUV4}) becomes $\cos\left(\frac{u_k-v_{i(k)}}{2}\right)=1$ for $k=1, \cdots, N-1$ 
leading 
\beq
u_k=v_{i(k)} \quad \mbox{for } k=1, \cdots, N-1. 
\label{u=v_2}
\eeq
Thus, we get 
\beq
u_1=\cdots =u_{N-1} = v_{i(1)}=\cdots =v_{i(N-1)}=\frac{\pi}{N},  
\eeq
however which is not consistent to the $k=N$ part of the eq. (\ref{eqUV4}): 
\beq
\cos\left[\frac12 (u_1+\cdots +u_{N-1}+v_{i(1)}+\cdots +v_{i(N-1)})+\frac{\pi}{2N}\right]
=\cos\frac{\pi}{2N}. 
\eeq

For the remaining case $\frac12(u_1+\cdots +u_{N-1} +v_{i(1)} +\cdots +v_{i(N-1)})=0$, 
we have 
\beq
u_1+v_{i(1)}=\cdots =u_{N-1}+v_{i(N-1)}=0 
\label{u+v_3}
\eeq
from (\ref{yoi}), 
and then (\ref{eqUV4}) becomes $\cos\left(\frac{u_k-v_{i(k)}}{2}\right)=1$ for $k=1, \cdots, N-1$ 
meaning 
\beq
u_k=v_{i(k)} \quad \mbox{for } k=1, \cdots, N-1. 
\label{u=v}
\eeq
Eqs. (\ref{u+v_3}) and (\ref{u=v}) lead  
\beq
u_1=\cdots =u_{N-1}=v_{i(1)}=\cdots =v_{i(N-1)}=0 
\eeq
to give $u_N=v_{i(N)}=0$. It is compatible to the possibility (\ref{yoi}). 

Therefore, the analysis shows that eqs. (\ref{eqUV4}) allow the solution 
$u_1=\cdots =u_N=v_1=\cdots =v_N=0$ only. 
We conclude that the equation (\ref{eqUV}) has the unique solution $U=V=1$ 
with choice of $r=\cot\varphi$ in the range $0 < \varphi \leq \frac{\pi}{2N}$.

\section{Explicit Form of the matrices $\gamma_{\mu}$, $P_{\mu}$, $P_{\mu}'$}
\label{sec:GP}
\setcounter{equation}{0}
\renewcommand{\theequation}{C.\arabic{equation}}

We present the explicit form of the matrices $\gamma_{\mu}$ and $P_{\mu}$ 
appearing in the naive lattice models of four-dimensional ${\cal N}=2,\,  4$ theories 
as well as $P'_{\mu}$ 
in their slightly modified versions. 
In what follows, $\sigma_1$, $\sigma_2$ and $\sigma_3$ represent the Pauli matrices. 

\subsection{Naive Model for four-dimensional ${\cal N}=2$ Theory}
First, in the naive model for ${\cal N}=2$, $\gamma_\mu$ is written as the form 
\beq
\gamma_\mu =   -i\left( \begin{array}{cc} 0 & \xi_\mu \\
                                \xi_\mu^T &   0  \end{array}\right), 
\label{gamma_mtrx}
\eeq
where 
\beq
 \xi_1 = \left(\begin{array}{cc}
           & \sigma_1 \\ 
 -i\sigma_2 &         \end{array}\right), \quad 
\xi_2 = \left(\begin{array}{cc}
            & -\sigma_3 \\ 
 {\bf 1}_2  &          \end{array}\right), \quad
\xi_3 = \left(\begin{array}{cc}
i\sigma_2  &          \\ 
            & \sigma_1 \end{array}\right), \quad
\xi_4 = \left(\begin{array}{cc}
 -{\bf 1}_2 &           \\ 
            & -\sigma_3  \end{array}\right). 
\label{gamma_naive_N=2}
\eeq

Also, 
$P_{\mu}$ are written as   
\beq
P_{\mu} = \left(\begin{array}{cc} 0 & \nu_{\mu} \\ 
                                  \nu_{\mu} & 0    \end{array}\right)
= \sigma_1\otimes \nu_{\mu} 
\quad \mbox{for  }\mu = 1, \cdots,  3, \qquad  
P_4 = \sigma_2\otimes {\bf 1}_4 
\eeq
with 
\beq
\nu_1 = \left(\begin{array}{cc} 0 & \sigma_2 \\ 
                                  \sigma_2 & 0    \end{array}\right), 
\quad \nu_2 = \left(\begin{array}{cc} 0 & -i{\bf 1}_2 \\ 
                                  i{\bf 1}_2 & 0    \end{array}\right), 
\quad \nu_3 = \left(\begin{array}{cc} -\sigma_2 & 0 \\ 
                                  0 & \sigma_2    \end{array}\right). 
\eeq

\subsection{Slightly Modified Model for four-dimensional ${\cal N}=2$ Theory}
In the slightly modified model for four-dimensional 
${\cal N}=2$ theory, the gamma matrices are same as (\ref{gamma_mtrx}), (\ref{gamma_naive_N=2}) 
in the naive model, while $P_{\mu}'$ are of the form 
\beq
P'_{\mu} = -i\left(\begin{array}{cc} 0 & \nu'_{\mu} \\ 
                                  -\nu_{\mu}^{\prime T} & 0    \end{array}\right)
\eeq
with 
\beq
\nu'_1 = \left(\begin{array}{cc} 0 & i\sigma_2 \\ 
                                  \sigma_1 & 0    \end{array}\right), 
\quad \nu'_2 = \left(\begin{array}{cc} 0 & {\bf 1}_2 \\ 
                                  -\sigma_3 & 0    \end{array}\right), 
\quad \nu'_3 = \left(\begin{array}{cc} -i\sigma_2 & 0 \\ 
                                  0 & \sigma_1    \end{array}\right), 
\quad \nu'_4 = {\bf 1}_4. 
\label{P'_N=2}
\eeq  

\subsection{Naive Model for four-dimensional ${\cal N}=4$ Theory}
$\gamma_{\mu}$ appearing in the naive model for ${\cal N}=4$ take the form of 
(\ref{gamma_mtrx}) with 
\bea
 & & \xi_1 = \left(\begin{array}{cccc}
       &         &        & \sigma_1 \\   
       &         & -\sigma_1 &       \\
       & \sigma_1&        &          \\
-i\sigma_2 &    &        &           \end{array}\right), \quad            
\xi_2 = \left(\begin{array}{cccc}
       &         &        & -\sigma_3 \\   
       &         & -\sigma_3 &       \\
       & \sigma_3&        &          \\
{\bf 1}_2 &    &        &           \end{array}\right), \quad            
\xi_3 = \left(\begin{array}{cccc}
       &         & \sigma_1&         \\   
       &         &         & \sigma_1 \\
 -\sigma_1&      &        &          \\
       & -i\sigma_2 &        &           \end{array}\right), \nn \\
 & & \xi_4 = \left(\begin{array}{cccc}
       &         & \sigma_3 &        \\   
       &         &          & -\sigma_3 \\
 -\sigma_3&         &        &          \\
       & {\bf 1}_2 &        &           \end{array}\right). 
\label{gamma_naive_N=4}       
\eea

$P_{\mu}$ are expressed as 
\beq
P_{\mu} = -i \left(\begin{array}{cc} 0 & \omega_{\mu} \\
                        -\omega_{\mu}^T & 0 \end{array} \right), 
\label{Pmu_N=4}
\eeq
where 
\bea
 & & \omega_1 = \left(\begin{array}{cccc}
        &           &          &  i\sigma_2  \\
        &           & \sigma_1 &             \\
        & \sigma_1  &          &             \\
-\sigma_1 &         &          &             \end{array} \right), \quad 
\omega_2 = \left(\begin{array}{cccc}
        &           &          &  {\bf 1}_2  \\
        &           & \sigma_3 &             \\
        & \sigma_3  &          &             \\
\sigma_3 &         &          &             \end{array} \right), \quad 
\omega_3 = \left(\begin{array}{cccc}
        &           & -\sigma_1 &            \\
        &           &           & i\sigma_2 \\
-\sigma_1&           &          &             \\
        & -\sigma_1 &          &             \end{array} \right), \nn \\
 & & 
\omega_4 = \left(\begin{array}{cccc}
        &           & -\sigma_3 &            \\
        &           &           & {\bf 1}_2 \\
-\sigma_3&           &          &             \\
        & \sigma_3  &          &             \end{array} \right).
\label{Pmu_N=4_4d}
\eea     

\subsection{Slightly Modified Model for four-dimensional ${\cal N}=4$ Theory}
For the slightly modified model in the case of ${\cal N}=4$ theory, 
$\gamma_{\mu}$ are given by (\ref{gamma_mtrx}), (\ref{gamma_naive_N=4}), and 
$P'_{\mu}$ are written as 
\beq
P'_{\mu} = -i \left(\begin{array}{cc} 0 & \omega'_{\mu} \\
                        -\omega_{\mu}^{\prime T} & 0 \end{array} \right), 
\label{Phat_N=4}
\eeq
where 
\bea
 & & \omega'_1 = \left(\begin{array}{cccc}
        &           &          &  i\sigma_2  \\
        &           & i\sigma_2 &             \\
        &-i\sigma_2  &          &             \\
-\sigma_1 &         &          &             \end{array} \right), \quad 
\omega'_2 = \left(\begin{array}{cccc}
        &           &           &  {\bf 1}_2  \\
        &           & {\bf 1}_2 &             \\
        & {\bf 1}_2 &          &             \\
\sigma_3 &         &          &             \end{array} \right), \quad 
\omega'_3 = \left(\begin{array}{cccc}
        &           & -\sigma_1 &            \\
        &           &           & \sigma_1 \\
-\sigma_1&           &          &             \\
        & i\sigma_2 &          &             \end{array} \right), \nn \\
 & & 
\omega'_4 = \left(\begin{array}{cccc}
        &           & -\sigma_3 &            \\
        &           &           & {\bf 1}_2 \\
-\sigma_3&           &          &             \\
        & \sigma_3 &          &             \end{array} \right).
\label{Phat_N=4_4d}
\eea



\begin{thebibliography}{999}

\bibitem{sugino}
F.~Sugino, 
{\em A lattice formulation of super Yang-Mills theories with exact
supersymmetry}, JHEP {\bf 0401} (2004) 015,  
[{\tt hep-lat/0311021}].
\bibitem{sugino2}
F.~Sugino, 
{\em Super Yang-Mills theories on the two-dimensional lattice with exact 
supersymmetry}, 
JHEP {\bf 0403} (2004) 067,  
[{\tt hep-lat/0401017}].

\bibitem{tft}
E.~Witten,
{\em Topological Quantum Field Theory}, 
Commun.\ Math.\ Phys.\  {\bf 117} (1988) 353; 
{\em Introduction To Cohomological Field Theories},
Int.\ J.\ Mod.\ Phys.\ A {\bf 6} (1991) 2775.

\bibitem{btft}
C.~Vafa and E.~Witten,
{\em A Strong coupling test of S duality}, 
Nucl.\ Phys.\ B {\bf 431} (1994) 3, 
[{\tt hep-th/9408074}].

R.~Dijkgraaf and G.~W.~Moore,
{\em Balanced topological field theories},
Commun.\ Math.\ Phys.\  {\bf 185} (1997) 411, 
[{\tt hep-th/9608169}].

M.~Blau and G.~Thompson,
{\em Aspects of $N(T) \geq 2$ topological gauge theories and D-branes}, 
Nucl.\ Phys.\ B {\bf 492} (1997) 545, 
[{\tt hep-th/9612143}].

J.~M.~F.~Labastida and C.~Lozano,
{\em Mathai-Quillen formulation of twisted N = 4 supersymmetric gauge 
theories in four dimensions}, 
Nucl.\ Phys.\ B {\bf 502} (1997) 741, 
[{\tt hep-th/9702106}].

\bibitem{catterall}
S.~Catterall, 
{\em Lattice supersymmetry and topological field theory},
JHEP {\bf 0305} (2003) 038, 
[{\tt hep-lat/0301028}].

S.~Catterall and S.~Ghadab, 
{\em Lattice sigma models with exact supersymmetry},
JHEP {\bf 0405} (2004) 044, 
[{\tt hep-lat/0311042}].

A.~D'Adda, I.~Kanamori, N.~Kawamoto and K.~Nagata, 
{\em Twisted superspace on a lattice}, 
[{\tt hep-lat/0406029}].

J.~Giedt and E.~Poppitz, 
{\em Lattice supersymmetry, superfields and renormalization}, 
JHEP {\bf 0409} (2004) 029, 
[{\tt hep-th/0407135}].

\bibitem{kikukawa-nakayama}
S.~Catterall and S.~Karamov, 
{\em Exact lattice supersymmetry: the two-dimensional N = 2 Wess-Zumino  
model},
Phys.\ Rev.\ D {\bf 65} (2002) 094501, 
[{\tt hep-lat/0108024}]; 
{\em A lattice study of the two-dimensional Wess Zumino model},
Phys.\ Rev.\ D {\bf 68} (2003) 014503, 
[{\tt hep-lat/0305002}].

Y.~Kikukawa and Y.~Nakayama, 
{\em Nicolai mapping vs. exact chiral symmetry on the lattice},
Phys.\ Rev.\ D {\bf 66} (2002) 094508, 
[{\tt hep-lat/0207013}].

K.~Fujikawa, 
{\em N = 2 Wess-Zumino model on the d = 2 Euclidean lattice},
Phys.\ Rev.\ D {\bf 66} (2002) 074510, 
[{\tt hep-lat/0208015}].

M.~Bonini and A.~Feo, 
{\em Wess-Zumino model with exact supersymmetry on the lattice}, 
JHEP {\bf 0409} (2004) 011, 
[{\tt hep-lat/0402034}].

A.~Kirchberg, J.~D.~Lange and A.~Wipf, 
{\em From the Dirac operator to Wess-Zumino models on spatial lattices}, 
[{\tt hep-th/0407207}].

\bibitem{ichimatsu}
K.~Itoh, M.~Kato, H.~Sawanaka, H.~So and N.~Ukita, 
{\em Novel approach to super Yang-Mills theory on lattice: Exact fermionic 
symmetry and 'Ichimatsu' pattern},
JHEP {\bf 0302} (2003) 033, 
[{\tt hep-lat/0210049}].

M.~Harada and S.~Pinsky,
{\em $N = (1,1)$ super Yang-Mills on a (2+1) dimensional transverse lattice  with 
one exact supersymmetry}, 
Phys.\ Lett.\ B {\bf 567} (2003) 277, 
[{\tt hep-lat/0303027}]; 
{\em A solution to the fermion doubling problem for supersymmetric theories on 
the transverse lattice}, 
[{\tt hep-lat/0408026}].

\bibitem{kaplan}
D.~B.~Kaplan, E.~Katz and M.~\"{U}nsal,
{\em Supersymmetry on a spatial lattice},
JHEP {\bf 0305} (2003) 037, 
[{\tt hep-lat/0206019}].

\bibitem{kaplan2}
A.~G.~Cohen, D.~B.~Kaplan, E.~Katz and M.~\"{U}nsal,
{\em Supersymmetry on a Euclidean spacetime lattice. I: A target theory with  
four supercharges},
JHEP {\bf 0308} (2003) 024, 
[{\tt hep-lat/0302017}];
{\em Supersymmetry on a Euclidean spacetime lattice. II: Target theories 
with eight supercharges}, 
JHEP {\bf 0312} (2003) 031,  
[{\tt hep-lat/0307012}].

\bibitem{kaplan_related}
J.~Nishimura, S.~J.~Rey and F.~Sugino,
{\em Supersymetry on the noncommutative lattice},
JHEP {\bf 0302} (2003) 032, 
[{\tt hep-lat/0301025}].

J.~Giedt, E.~Poppitz and M.~Rozali,
{\em Deconstruction, lattice supersymmetry, anomalies and branes},
JHEP {\bf 0303} (2003) 035, 
[{\tt hep-th/0301048}].

M.~\"{U}nsal,
{\em Regularization of non-commutative SYM by orbifolds with discrete torsion and
SL(2,Z) duality}, 
[{\tt hep-th/0409106}].

\bibitem{review}
I.~Montvay, 
{\em Supersymmetric Yang-Mills theory on the lattice},
Int.\ J.\ Mod.\ Phys.\ A {\bf 17} (2002) 2377, 
[{\tt hep-lat/0112007}].

A.~Feo, 
{\em Predictions and recent results in SUSY on the lattice}, 
[{\tt hep-lat/0410012}].

\bibitem{elitzur}
S.~Elitzur, E.~Rabinovici and A.~Schwimmer, 
{\em Supersymmetric Models On The Lattice},
Phys.\ Lett.\ B {\bf 119} (1982) 165.



\bibitem{luscher} 
M.~L\"{u}scher, 
{\em Topology and the axial anomaly in abelian lattice gauge theories}, 
Nucl.\ Phys.\ B {\bf 538} (1999) 515, [{\tt hep-lat/9808021}]. 

T.~Fujiwara, H.~Suzuki and K.~Wu, 
{\em Axial anomaly in lattice Abelian gauge theory in arbitrary dimensions}, 
Phys.\ Lett.\ B {\bf 463} (1999) 63, [{\tt hep-lat/9906016}]; 
{\em Non-commutative differential calculus and the axial anomaly in Abelian 
lattice gauge theories}, 
Nucl.\ Phys.\ B {\bf 569} (2000) 643, [{\tt hep-lat/9906015}].

\bibitem{seiberg-witten}
N.~Seiberg and E.~Witten, 
{\em Electric - magnetic duality, monopole condensation, and confinement in 
N=2 supersymmetric Yang-Mills theory},
Nucl.\ Phys.\ B {\bf 426} (1994) 19
[Erratum-ibid.\ B {\bf 430} (1994) 485], 
[{\tt hep-th/9407087}]. 
\bibitem{giedt}
J.~Giedt, 
{\em Non-positive fermion determinants in lattice supersymmetry},
Nucl.\ Phys.\ B {\bf 668} (2003) 138, 
[{\tt hep-lat/0304006}];
{\em The fermion determinant in (4,4) 2d lattice super-Yang-Mills},
Nucl.\ Phys.\ B {\bf 674} (2003) 259, 
[{\tt hep-lat/0307024}]; 
{\em Deconstruction, 2d lattice Yang-Mills, and the dynamical lattice
spacing}, 
[{\tt hep-lat/0312020}].

\bibitem{sugihara}
T.~Sugihara, 
{\em Density matrix renormalization group in a two-dimensional $\lambda \phi^4$ 
Hamiltonian lattice model}, 
JHEP {\bf 0405} (2004) 007, 
[{\tt hep-lat/0403008}].

\bibitem{mandelstam}
S.~Mandelstam, 
{\em Light Cone Superspace And The Ultraviolet Finiteness Of The N=4 
Model}, 
Nucl.\ Phys.\ B {\bf 213} (1983) 149.


\bibitem{DVVBS}
R.~Dijkgraaf, E.~Verlinde and H.~Verlinde, 
{\em Matrix string theory}, 
Nucl.\ Phys.\ B {\bf 500} (1997) 43, [{\tt hep-th/9703030}].

T.~Banks and N.~Seiberg, 
{\em Strings from matrices}, 
Nucl.\ Phys.\ B {\bf 497} (1997) 41, [{\tt hep-th/9702187}].



\end{thebibliography}
\end{document}